\documentclass[lectures]{pcms-l}

\usepackage{amssymb}

\usepackage{graphicx}

\numberwithin{equation}{chapter}

\begin{document}


\frontmatter
\tableofcontents


\mainmatter


\LectureSeries[Statistical Mechanics]
{Three lectures on statistical mechanics \author{Veit Elser}}



\address{Department of Physics, Cornell University}
\email{ve10@cornell.edu}






%
\lecture{Mechanical foundations}

\section*{Laws, hypotheses, and models}

It is widely held that thermodynamics is built upon a foundation of laws that go beyond the standard laws of mechanics. This is not true. Even the microscopic level of thermodynamic description, called statistical mechanics, requires only the application of strict mechanical principles. The ``laws" that appear to be new and indispensable to statistical mechanics are really \textit{hypotheses} about the mathematical consequences of the laws of mechanics. One such consequence, the \textit{ergodic hypothesis}, has been rigorously demonstrated only for some simple model systems. However, its validity is not in doubt for the broad range of systems where it applies, that it is treated as an actual law, perhaps best known as the principle of equal \textit{a priori} probability. We should really think of this law as a \textit{model} for the statistical properties of mechanical systems. This model need not always apply, but when it does it is tremendously useful. Rather than define this hypothesis/model in the abstract, we illustrate it with a simple example.

\section*{Soft billiards}

Consider two identical mass $m$ particles moving in a two-dimensional world and interacting by a potential that only depends on the distance between their centers. Because the ergodic hypothesis only applies to bounded systems we place the particles in a box. To minimize the effects of the shape of the box, we let the world be a flat torus, in other words, a square with opposite edges identified.

The mechanical description of this system is simplified by working not with the positions $\mathbf{r}_1$ and $\mathbf{r}_2$ of the particles, but instead, their centroid $\mathbf{R}=(\mathbf{r}_1+\mathbf{r}_2)/2$ and relative position $\mathbf{r}=\mathbf{r}_1-\mathbf{r}_2$. The equations of motion for $\mathbf{R}$ and $\mathbf{r}$ are independent, the equations for $\mathbf{R}$ being that of a particle of mass $2m$ subject to no forces. On the other hand, the equations for $\mathbf{r}$ are more interesting and describe a single particle of ``reduced" mass $\mu=m/2$ subject to a potential $U(\mathbf{r})$ fixed in the torus. We will consider an especially simple potential that takes only two values: $U(\mathbf{r})=U_0$ when $|\mathbf{r}|<b$, and $U(\mathbf{r})=0$ otherwise. As shown in Figure 1, we keep the interaction range $b$ smaller than half the edge of the square, and place the origin $\mathbf{r}=0$ at the center.

Figure 1 shows two renderings of the motion of the relative position $\mathbf{r}$ : as a trajectory in the torus with one circular obstacle, or as a trajectory through a periodic crystal of obstacles. The nature of the motion depends both on the sign of $U_0$ and the total energy $E$ of the equivalent single-particle. Shown is the case $0<U_0<E$, such that the particle has sufficient energy to find its way to any place on the torus. If instead we had $0<E<U_0$, the particle would be excluded from the interior of the circular obstacle; this is the ``hard billiards" model studied by Yakov Sinai \cite{Sinai}. We will examine the ``soft billiards" case, $0<U_0<E$, because it better demonstrates the scope of the ergodic hypothesis.

\begin{figure}
\includegraphics{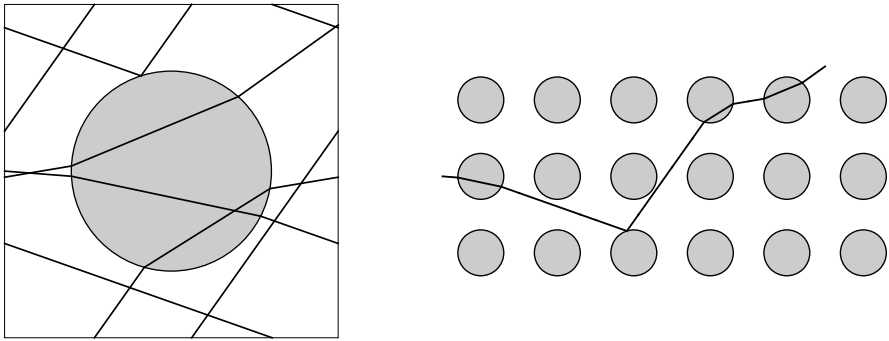}
\caption{Two renderings of a trajectory in soft-billiards: on the torus (left panel), and in a periodic crystal (right panel). The shaded circular regions are at higher potential energy $U_0$ relative to zero potential energy elsewhere.}
\label{}
\end{figure}

In the soft billiards model, the trajectory is comprised of linear pieces that undergo sudden deflections when encountering the circular region. The principle that determines the deflection angle is the same as the principle behind Snell's Law: conservation of energy, and, conservation of momentum tangential to the boundary. Upon entering the circular higher-potential region, the particle's momentum perpendicular to the boundary abruptly decreases in order to have the requisite lower kinetic energy, while the tangential momentum is unchanged. Had we made our potential $U(\mathbf{r})$ rise continuously from $0$ to $U_0$, the momentum-kick would be spread over a finite range; however, regularizing the motion in these short intervals has no effect on the highly chaotic character of the trajectory on larger scales. At sufficiently glancing incidence on the circle, energy and momentum conservation forbid the trajectory to cross into the interior. In this case the particle is simply reflected by the boundary, exactly as in the hard billiards model.

\begin{figure}
\includegraphics{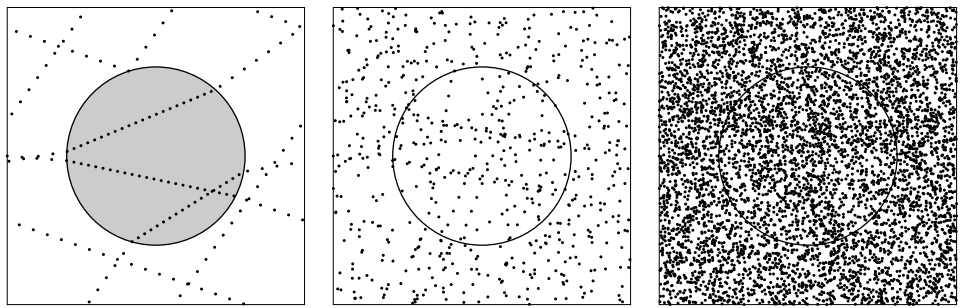}
\caption{\textit{Left:} Stroboscopic rendering of the trajectory in Figure 1. The particle has lower kinetic energy in the circular region and moves slower there. Deflections, caused by the momentum-kicks at the potential discontinuity, are analogous to Snell's law in optics. \textit{Middle:} Same as the image on the left, but with tenfold increase in time span (also lower stroboscopic rate).  \textit{Right:} Tenfold increase in time span over middle image.}
\label{}
\end{figure}

Because of chaos, there is not much point in trying to predict where the particle will be at any time. But also because of chaos, there are a number of questions worth asking that we would not ask in the absence of chaos. Since it seems plausible that the particle will visit every part of the torus eventually, we can ask with what frequency the different parts are visited. Will every part inside the circle be visited with the same frequency, and so too every part outside, but with some other frequency? Is the frequency in inverse proportion to the speed of the particle in the region, so that the particle spends more time in the region of higher potential energy?

The numerical experiment shown in Figure 2 suggests we were correct in conjecturing uniform rates of visitation. On the other hand, the evidence points to equal rates in the two regions, contrary to our intuition. Before we see how the ergodic hypothesis can explain both of these observations, we need to take a step back and recall how our model is described in the Hamiltonian formalism of mechanics.

Our model has two degrees of freedom corresponding to the $x$ and $y$ coordinate of the particle\footnote{These have the topology of angular variables for our periodic boundary conditions.} . Associated with each of these is a conjugate momentum, $p_x$ and $p_y$, and the combined space of coordinates and momenta is called \textit{phase space}. The equations of motion follow from the Hamiltonian, which in our case is
\begin{equation*}
H(x,y,p_x,p_y)=\frac{p_x^2}{2\mu}+\frac{p_y^2}{2\mu}+U(x,y),
\end{equation*}
\begin{equation*}
U(x,y)=\left\{
\begin{array}{rl}
U_0,&x^2+y^2<b^2\\
0,&\mbox{otherwise.}
\end{array}\right.
\end{equation*}

Given some property $\theta(x,y,p_x,p_y)$ that depends on the Hamiltonian variables we can form averages in the sense of dynamics as follows:
\begin{equation}\label{timeave}
\overline{\theta}=\frac{1}{T}\int_0^T dt\; \theta(x(t),y(t),p_x(t),p_y(t)),
\end{equation}
where the time evolution of the variables is determined by the Hamiltonian and a choice of initial conditions. The weak form of the ergodic hypothesis asserts that in the limit of large $T$ any such dynamical average can alternatively be computed as a phase space average with respect to some distribution $\rho$,
\begin{equation*}
\langle\theta\rangle=\int dx\, dy\, dp_x\, dp_y\; \rho(x,y,p_x,p_y)\, \theta(x,y,p_x,p_y),
\end{equation*}
thus sidestepping the intractability of chaotic time evolution.

The ergodic hypothesis does not have much value to physicists unless the distribution $\rho$ is known rather precisely. Remarkably, there is a single very simple distribution that is believed to apply to almost all bounded systems whose dynamics is sufficiently chaotic. To formulate this distribution we need to acknowledge that even in the presence of chaos there will be conserved quantities that the distribution must respect. Typically, as in our example once we  eliminated the trivial centroid motion, only the energy is conserved. The strong form of the ergodic hypothesis, or how the hypothesis is usually interpreted by physicists, asserts that $\rho$ is supported on a connected level set of the conserved quantities and is otherwise uniform. In the case of energy -- the Hamiltonian -- as the only conserved quantity, we then have
\begin{equation}\label{uniform}
\rho(x,y,p_x,p_y)=\rho_0\; \delta(H(x,y,p_x,p_y) - E),
\end{equation}
where $E$ is the value of the conserved energy, $\delta$ is the Dirac delta-function, and $\rho_0$ is a normalization constant. The statement
\begin{equation}\label{ergodichyp}
\overline{\theta}=\langle\theta\rangle,
\end{equation}
with $\rho$ in the phase space average given by (\ref{uniform}), is how we will interpret the ergodic hypothesis from now on. A more descriptive term for the same thing is the principle of \textit{equal a priori probability}: over the course of time, the system visits all states consistent with the conserved quantities with equal frequency.

We are now ready to apply the ergodic hypothesis to find the frequency our particle visits a particular point $(x_0,y_0)$ on the torus. For this we calculate the phase space average of
\begin{equation*}
\theta=\delta(x-x_0)\delta(y-y_0),
\end{equation*}
and find,
\begin{align*}
\langle\theta\rangle&=\rho_0\int_{-\infty}^\infty dp_x \int_{-\infty}^\infty dp_y\; \delta(H(x_0,y_0,p_x,p_y) - E)\\
&=\rho_0\int_0^\infty 2\pi p dp\; \delta(p^2/(2\mu)-K(x_0,y_0)),
\end{align*}
where we have transformed to polar coordinates in the momentum integrals and defined the local kinetic energy
\begin{equation*}
K(x_0,y_0)=E-U(x_0,y_0).
\end{equation*}
Performing the final radial integral we obtain
\begin{equation*}
\langle\theta\rangle=2\pi\rho_0\mu,
\end{equation*}
provided $K(x_0,y_0)>0$. This result is remarkable mostly because it is independent of the point $(x_0,y_0)$, in agreement with the counter-intuitive findings of our numerical experiment.

Before we jump to the conclusion that the uniform sampling of position is a general feature of chaotic systems, we should calculate the corresponding phase space average for a soft billiards system in a dimension other than $d=2$. The result we obtain,
\begin{equation*}
\langle\theta\rangle\propto K(x_0,y_0,\ldots)^{d/2-1},
\end{equation*}
shows that $d=2$ is special for its sampling uniformity. On the other hand, the conclusion that in dimensions $d>2$ the dynamics samples points at a higher rate, where its kinetic energy (speed) is greater, is even more at odds with our intuition!

\section*{Quantum states}

Physicists have a much more explicit concept of the ``states" in a mechanical system than the phase-space formulation of the ergodic principle suggests in the abstract. That's because the true laws of mechanics are the laws of quantum mechanics, where the states of a closed system are discrete and can be counted. 

Statistical mechanics is usually understood to be applicable only when a system has very many degrees of freedom. But as we saw in the soft billiards system, this is not a requirement at all. However, we now show that this characterization is correct after all, if ``degrees of freedom" are reinterpreted as quantum mechanical states.

Consider a rectangular region of extent $[x_0,x_0+\Delta x]\times [y_0,y_0+\Delta y]$ in the soft billiards system, and assume the potential $U(x,y)$ is constant in this region. The solutions of the Schr\"odinger equation for a particle confined to this region are products of sinusoids,
\begin{equation*}
\Psi(x,y)=\sin{((x-x_0)p_x/\hbar)}\;\sin{((y-y_0)p_y/\hbar)},
\end{equation*}
where $\hbar$ is Planck's constant divided by $2\pi$ and the momenta are required to have discrete values determined by the vanishing boundary condition (on the rectangular region of interest):
\begin{align*}
\Delta x\; p_x /\hbar&=\pi n_x,\qquad n_x=1,2,3,\ldots\\
\Delta y\; p_y /\hbar&=\pi n_y,\qquad n_y=1,2,3,\ldots
\end{align*}
Classical mechanics is the $\hbar\to 0$ asymptotic limit of quantum mechanics. For any range of the momenta, say $dp_x$ and $dp_y$, that we might care to resolve in classical mechanics, there are very many quantum states (allowed values of $n_x$ and $n_y$):
\begin{equation*}
\left(\frac{\Delta x\; dp_x}{h/2}\right)\left(\frac{\Delta y\; dp_y}{h/2}\right).
\end{equation*}
Given a similarly finite extent to which the kinetic energy can be resolved,
\begin{equation*}
K<\frac{p^2}{2\mu}<K+dK,
\end{equation*}
we obtain the total number of states as the product of the momentum-space state density, $4\Delta x\Delta y/h^2$, and the area of a momentum-space annulus in the positive quadrant of radius $p=\sqrt{2\mu K}$ and width $dp=(\mu/p)dK$ :
\begin{equation}\label{dN}
dN\sim\left(\frac{4\Delta x\Delta y}{h^2}\right)\left(\frac{2\pi p}{4}\right) \left(\frac{\mu}{p}dK\right)\propto \Delta x\Delta y\; dK.
\end{equation}
The number of quantum states is thus proportional to the area of the rectangle, and independent of the kinetic energy (location of the rectangle in the billiards). A similar calculation, for states in a rectilinear region of billiards in $d$ dimensions, gives the result
\begin{equation}\label{rectdN}
dN\propto V K^{d/2-1} dK,
\end{equation}
where $V$ is the volume of the region. Both of these results are in complete agreement with the phase space averaging we performed earlier, which made no reference to quantum states.

Formula \eqref{rectdN} has a nice generalization, called \textit{Weyl's law} \cite{Weyl}, where the rectilinear region is replaced by a region of arbitrary shape in a Riemannian manifold. This includes the classical hard billiards problem: a particle confined inside a region of uniform potential. Choosing the potential inside equal to zero so the kinetic energy equals the total energy $E$, we write this generalization as
\begin{equation*}
dN\sim c\, V E^{d/2-1} dE,
\end{equation*}
where $c$ involves the mass of the particle, Planck's constant, the dimension, but no other characteristics of the region. The significance of this asymptotic result should not be overlooked, since not only is calculating the chaotic billiards trajectory of a classical particle intractable, so too is calculating the quantum spectrum of energy levels in an arbitrary region. In the next section we will need the integrated form, an asymptotic level counting formula:
\begin{equation}\label{levelcount}
N(E)\propto V E^{d/2}.
\end{equation}
The constant of proportionality is omitted, as it only depends on fixed constants. This is just the leading term of an asymptotic series, but the only one that matters in the classical limit. The correction terms depend on characteristics of the billiards region other than its volume, and are the subject of Mark Kac's classic article \cite{Kac} on being able to ``hear the shape of a drum".

\section*{Kinetic elasticity}

There is probably no better example of how quantum states assert themselves in classical mechanics than the kinetic polymer model to which we now turn. Figure 3 shows one configuration of the polymer: equal length rigid struts, free to pivot in the plane about joints which carry all the mass. This is also a ``microscopic" model in the sense that there are no phenomenological forces at a smaller scale, such as the friction we would expect in a macroscopic joint. For the sake of tractability, we allow the masses and struts to behave as phantom entities that may freely pass through each other. Apart from the latter, this is a reasonably realistic model of a polymer in vacuum, as bond-angle forces --- completely absent here --- are often significantly weaker than the forces that fix the bond lengths. We will add the effects of a solvent environment to the model in a future lecture.

\begin{figure}
\includegraphics{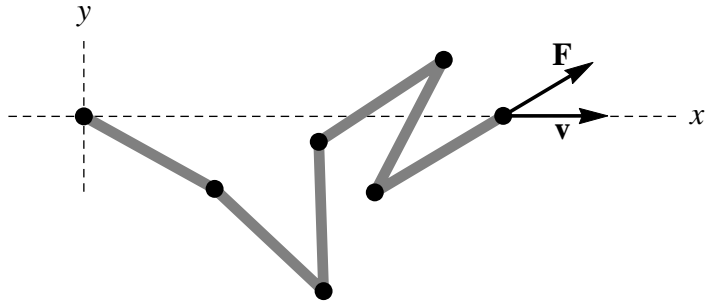}
\caption{A configuration of the freely-jointed polymer model with $n=6$ rigid struts. The polymer-end on the left is fixed to the origin while the end on the right is moved with constant velocity $\mathbf{v}$ by an external mechanism. The fluctuating force $\mathbf{F}$ acting on the external mechanism is parallel to the end strut.}
\label{}
\end{figure}

The quantum correspondence is seen in the behavior of the model when subject to an \textit{adiabatic process}. We will examine the process where an external mechanism fixes one end of the polymer and moves the other with a constant velocity. A polymer of $n$ struts therefore has $n-2$ degrees of freedom. We can think of the moving end  as an infinite (macroscopic) mass whose velocity $\mathbf{v}$ is unaffected by the motion of the polymer. The last strut of the polymer exerts a time-dependent force $\mathbf{F}(t)$ on the macroscopic mass (parallel to the strut). The polymer thus delivers instantaneous power $p(t)=\mathbf{F}(t)\cdot\mathbf{v}$ to the macroscopic mass, which is the negative of the power delivered to the polymer by the mass. Figure 4 shows a time series of $p(t)$ for a 6-strut polymer during a period of time where the moving end has moved only a small fraction of one strut-length. The erratic nature of this function reflects, of course, the highly chaotic dynamics of the polymer.

\begin{figure}
\includegraphics{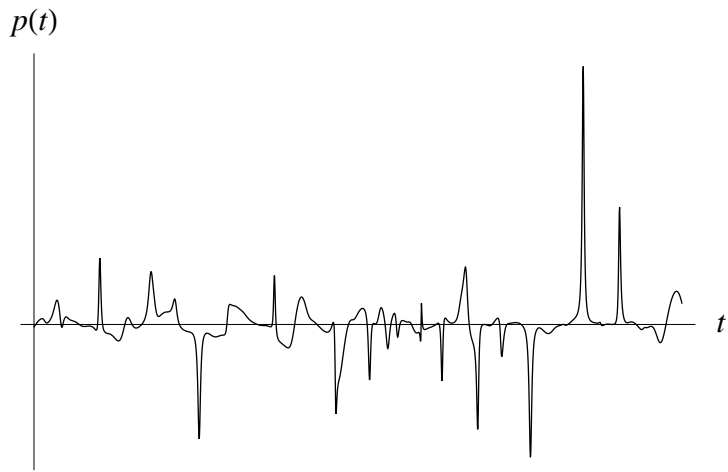}
\caption{Time series of the instantaneous power, $p(t)=\mathbf{F}(t)\cdot\mathbf{v}$, delivered by the polymer to the external mechanism.}
\label{}
\end{figure}

 We will perform a number of numerical experiments with the 6-strut polymer. In reporting the results, we use the strut length for our unit of length. The adiabatic process is begun with zero separation of the ends and randomly generated initial conditions consistent with one end fixed and the other moving with velocity $\mathbf{v}$. Our unit of speed is the root-mean-square speed of the masses at the start of the process. Likewise, our energy unit is the total energy --- entirely kinetic --- of the polymer at the start. The instantaneous energy of the polymer changes as a result of the fluctuating power:
 \begin{align}\label{deltaE}
 \Delta E&=-\int_0^t \mathbf{F}(t')\cdot\mathbf{v}\, dt'\\
 &=-\int_{\mathbf{r}(0)}^{\mathbf{r}(t)} \mathbf{F}(t')\cdot d\mathbf{r}.\notag
 \end{align}
 What makes a process \textit{adiabatic} is slowness in the changes of the external parameters. In the adiabatic limit the position of the moving end $\mathbf{r}(t)$ has hardly changed over a span of time during which many fluctuations in $\mathbf{F}(t)$ have occurred. This motivates us to rewrite \eqref{deltaE} as
 \begin{equation*}
 E(x)-E(0)=-\int_0^x F(x') dx',
 \end{equation*}
 where $x=|\mathbf{v}|dt = v t$ is the separation of the ends, and a time average defines a position-dependent force:
 \begin{equation*}
 F(x) dx = \overline{\mathbf{F}(t)}\cdot d\mathbf{r}.
 \end{equation*}
 Our numerical experiments are strictly mechanical: we do not perform any actual averages. Instead, we record the instantaneous energy $E$ of the polymer, at time $t$ and end-to-end distance $x=v t$, as the function $E(x)$. This function will have random features that vary from one set of initial conditions to another.

Figure 5 shows four $E(x)$ curves obtained when the moving end has speed $v=0.1$. While there is much randomness, there is also a clear trend: pulling on the polymer tends to \textit{increase} its energy. In Figure 6 we see the results of four more experiments, but with $v=0.01$. Two things have changed at the slower speed: fluctuations have been suppressed by about a factor of 10, the same as the reduction in speed, and there is better evidence of a well defined average energy. Superimposed on these curves is the simple function $\exp{(x^2/12)}$, whose significance will be made clear below.

\begin{figure}
\includegraphics{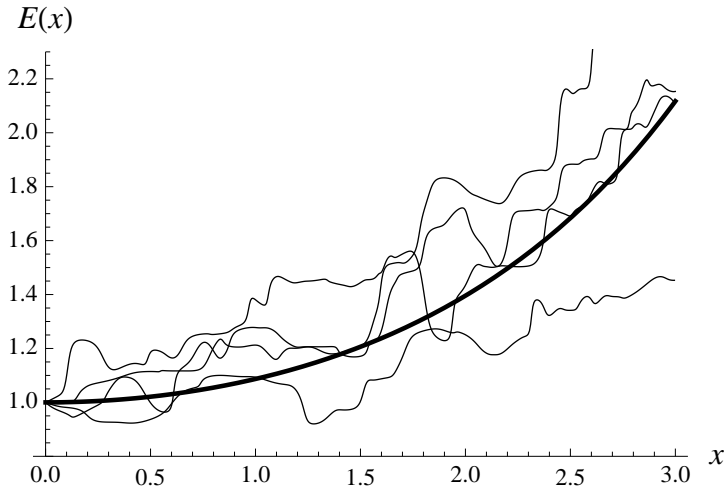}
\caption{Four energy vs. extension curves, $E(x)$, for the $n=6$ polymer model when the speed of the moving end is one-tenth the root-mean-square speed of the masses at $x=0$. The thick curve is the function $\exp{(x^2/12)}$.}
\label{}
\end{figure}

Since the equations of mechanics do not distinguish between the future and past directions of time, we know exactly what will happen in an experiment where the ends of the polymer are brought together from a stretched state: the energy will decrease along the same curve we found when it was stretched. If the mass at the end of the last strut was made finite, so its velocity could change, it would respond (as a new degree of freedom) to the rest of the polymer much as a mass attached to an elastic spring. From the quadratic behavior of $E(x)$ at small $x$ we see that this spring, at small extension, has a linear force law just as a conventional spring. What is curious about this particular spring, however, is the absence of any potential terms in its microscopic Hamiltonian for the storage of energy.

\begin{figure}
\includegraphics{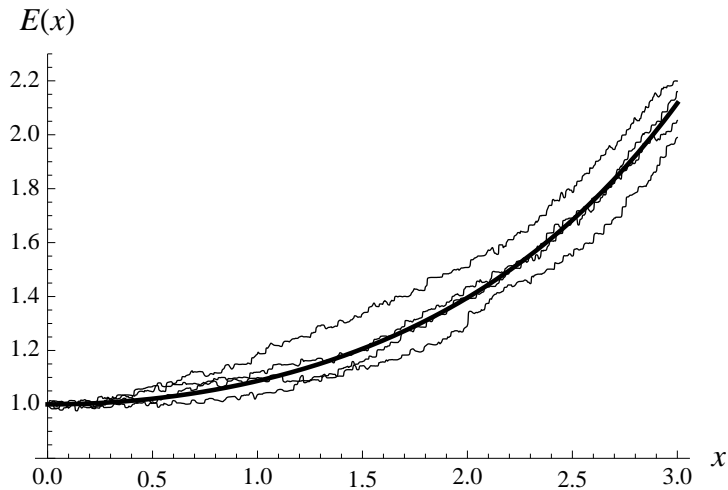}
\caption{Same as Figure 5, but with the rate of extension slower by a factor of 10.}
\label{}
\end{figure}

Although classical Hamiltonian mechanics does address the effects of adiabatic change when it is imposed on \textit{periodic} motion, for our chaotic polymer a more powerful tool is needed. That tool is the very simple behavior of a quantum system in an adiabatic process. The rule for quantum systems is that a state will evolve so as to remain in an instantaneous energy eigenstate even when a parameter $x$ in the (quantum) Hamiltonian is varied, provided the change is slow. The energy of the quantum system thus varies exactly as the energy $E(x)$ of a particular energy eigenstate. Intractability or ``non-integrability" of the classical equations of motion is, ironically, good in this respect because it means the corresponding quantum energy levels do not cross when a parameter is changed. The quantum analog of our polymer system, when prepared in the $N$th energy state for $x=0$, would still be in the $N$th level when $x$ is slowly changed to some other value.

The energy levels of the quantum analog of our polymer model are within our reach once we realize this model is a multi-dimensional billiards for which Weyl's asymptotic result \eqref{levelcount} applies, where $d=n-2$ is the number of degrees of freedom. The only remaining hurdle is determining $V(x)$, the dependence of the configuration space volume on the polymer's end-to-end separation $x$. Putting that aside for now, from \eqref{levelcount} we obtain
\begin{equation*}
E(N,x)\propto (N/V(x))^{2/(n-2)},
\end{equation*}
and since $N$ is constant in an adiabatic process,
\begin{equation*}
E(x)=E(0)\left(\frac{V(0)}{V(x)}\right)^{2/(n-2)}.
\end{equation*}

The calculation of $V(x)$ is a much studied problem in statistics. Parametrizing the polymer by the angles $\theta_1,\ldots,\theta_n$ of the struts relative to the axis of extension, an explicit formula takes the form
\begin{equation*}
V(x)=\int_0^{2\pi} d\theta_1\cdots \int_0^{2\pi} d\theta_n\, \delta(\cos{\theta_1}+\cdots+\cos{\theta_n}-x)\delta(\sin{\theta_1}+\cdots+\sin{\theta_n}).
\end{equation*}
Up to normalization, $V(x)$ is just the probability a random walk of $n$ unit steps arrives at a particular point, whose distance from the origin is $x$. For large $n$ and $x$ growing at most as $\sqrt{n}$ we can use the central limit theorem to get the estimate
\begin{equation}\label{centlim}
V(x)\propto \exp{(-x^2/n)},
\end{equation}
and thus
\begin{equation*}
E(x)=E(0)\exp{\left(\frac{2}{n(n-2)}\,x^2\right)}.
\end{equation*}
Central limit convergence is surprisingly rapid, as the comparison of this formula with the $n=6$ numerical experiment in Figure 6 shows.

\section*{Statistical equilibrium}

In mechanics we use the term \textit{equilibrium} to describe motion at its very simplest, where all the variables are time independent. Statistical mechanics has its own notion of equilibrium, and time is again at the center of the definition.

Time is the relevant quantity, both for the ergodic hypothesis and for a process to be adiabatic. A phase space average, by definition, is time independent. But it estimates the time averages that arise in our study of physical phenomena only when those averaging times are sufficiently large. In the soft billiards model ``sufficient" is translated to ``many deflections of the particle by the potential have occurred".

Adiabatic processes transform a system reversibly from one set of parameters to another set. But again, this is true only when the process is carried out over a long enough time. The function $E(x)$ for the energy of the polymer model, and its derivative giving the elastic force, is only well defined when the rate of extension is slow. The long time-average in this case is that of the fluctuating force generated by the polymer and acting on the external mechanism. Only when diverse polymer configurations are sampled in the time taken for $x$ to move through a small range does the force-average correspond to a true position-dependent force.

Equilibrium, in statistical mechanics, is not so much a state or behavior that a system might settle into, but a statement about observations or processes being carried out at the appropriate temporal scale. The conditions for equilibrium are also what make the subject difficult. In extreme but by no means exotic cases, the ergodic hypothesis is known to fail, and it might not be possible to exercise adiabatic control.

Consider the soft billiards model but with an attracting circular potential and negative energy: $U_0<E<0$. For these conditions the system is equivalent to  hard \textit{circular} billiards, an ``integrable" model because the trajectory (an infinite stellated polygon) is not very sensitive to initial conditions. The ergodic hypothesis fails because a circular set of positions (depending on initial conditions) is never reached. We can ask if this state-of-affairs is robust with respect to perturbations. Unfortunately, the breakdown of the ergodic hypothesis persists even for a finite range of smooth perturbations of the circular shape \cite{Berry}. This is consistent with the statement of the Kolmogorov-Arnold-Moser (KAM) theorem\footnote{The KAM theorem applies to Hamiltonians with smooth potential functions, and therefore not to billiards.}. On the other hand, an arbitrarily small protrusion on the wall of this integrable billiard can be sufficiently randomizing to restore the ergodic hypothesis. The time scale for ``statistical equilibrium", by construction, can thereby be made arbitrarily long. 

Long time scales were also in evidence in our polymer model experiments. The long undulations that differentiate the four curves in Figure 6 suggest that the polymer has very slow modes of oscillation whose period is still beyond the time scale of the extension. Slow modes are a general feature of systems with many degrees of freedom; they represent the main mechanism whereby a system can be said to be ``out of equilibrium".

Slow modes and results such as the KAM theorem are what dim the hopes of proving a general form of the strong ergodic hypothesis. Statistical mechanics, in response, has adopted the ``law" of equal \textit{a priori} probabilities as a model, with the understanding that this model may not apply to all systems.

\section*{Problems for study}
\subsection{Snell's law for particles}
A mass $m$ particle in two dimensions moves in a potential that has only two values, as in the soft billiards system. Its speed therefore takes two values: a high speed $v_1$ and a low speed $v_2$. Suppose the particle starts in the high speed region and is incident on a region of high potential with straight boundary, where its speed will be slow. Using momentum conservation, only in the force-free direction parallel to the boundary, show that the angles of incidence of the particle satisfy $v_1\sin{\theta_1}=v_2\sin{\theta_2}$. Here $\theta_1$ and $\theta_2$ are the angles subtended by the trajectory and the perpendicular to the noundary; $\theta_1=\theta_2=0$ corresponds to normal incidence and no deflection.

When $\theta_1$ exceeds a particular value this Snell's law for particles has no solution and the particle is reflected back into the high speed region. Find this maximum angle and determine the law of reflection, again using momentum conservation.

\subsection{Tracer particle analysis of soft billiards}
The ergodic hypothesis is difficult to prove, even for simple systems. This exercise should at least make the hypothesis plausible for the soft billiards system.

Instead of following a single very complex trajectory, we will analyze simple families of trajectories over a limited time. The family we have in mind is best described as a set of tracer particles initially arranged with uniform density $\rho$ along the $y$-axis. All particles are in the low potential region and have speed $v_1$ and velocity in the direction of the positive $x$-axis.

First show that the total time spent by all the tracer particles crossing a circular region of radius $b$ in the low potential region is
\begin{equation*}
T_1=(\rho/v1)\pi b^2.
\end{equation*}

Next suppose the parallel streaming tracer particles encounter a circular region of high potential, where their speed slows to $v_2$. The time spent by all the tracer particles crossing this region can be written as
\begin{equation*}
T_2=\int_{-b}^{+b} (\rho dy) T(y),
\end{equation*}
where $T(y)$ is the time spent in the region by a tracer particle with initial offset $y$ from the center of the circle. Use Snell's law to show that
\begin{equation*}
T(y)=\left\{
\begin{array}{cl}
(2b/v_2)\sqrt{1-(v_1/v_2)^2 (y/b)^2}\;,& |y|<(v_2/v_1)b\\
0\;,& \mbox{otherwise}.
\end{array}\right.
\end{equation*}
Now evaluate the integral for $T_2$ and observe that it exactly equals $T_1$.

\subsection{Weyl's law}
Derive \eqref{rectdN} by repeating the calculation of the result \eqref{dN} for a billiards in $d$ dimensions.

\subsection{Central limit analysis of polymer}
Apply the central limit theorem to a random walk of $n$ unit steps, isotropically distributed in two dimensions, to derive the large $n$ estimate \eqref{centlim}.

\subsection{Slow modes of a polymer}
The slowest mechanical modes of a system bound the time scale of adiabatic processes. In a gas the slow modes are sound, the slowest being coherent oscillatory motion on large scales, where the density in one half grows at the expense of the other half. Speculate what might be the slowest modes in the kinetic polymer system.

\lecture{Temperature and entropy}

\section*{The thermodynamic limit}

In the first lecture we saw that statistical mechanics provides a quantitative description of mechanical phenomena where time has been completely eliminated. This is a great benefit in systems whose dynamics is chaotic, and conversely, the statistical description very much relies on chaos for the foundational model to be valid. This part of the subject applies just as much to systems of few degrees of freedom as it does to systems with many. Another level of modeling applies when systems become very large, the regime of phenomena in the \textit{thermodynamic limit}. As the name suggests, it is only in this limit that the concept of \textit{temperature} makes sense. It is also in this thermodynamic limit, of systems with a well defined temperature, that \textit{entropy} may be defined as a commodity that is interchangeable with energy.

\section*{A model for the number of states}

An interesting mathematical quantity we can define for an arbitrary mechanical system (for which the ergodic hypothesis holds) is the \textit{number-of-states} function:
\begin{equation}\label{omega}
\Omega(E)=\int dq_1\cdots dp_1\cdots\;\delta(H(q_1,\ldots,p_1,\ldots)-E).
\end{equation}
This is the phase space integral of the uniform distribution promised by the ergodic hypothesis. With proper normalization its value is unity; without this normalization the integral $\Omega(E)$ is interpreted as the number of states of the system with energy in a fixed range of arbitrarily narrow width about $E$.

To get our bearings we will calculate the number-of-states function for a system of $n$ identical and weakly interacting particles in a three dimensional box of volume $V$: the ideal gas. In the limit of weak interactions the energy is just the kinetic energy of the particles. However, the interactions cannot be switched off completely because then each particle's energy is fixed by the initial conditions, contrary to the ergodic hypothesis. Ignoring the ergodicity restoring interactions in $H$, the position integrals in \eqref{omega} give a factor $V$ for each particle, and the $3n$ momentum integrals give the volume of a spherical shell in $3n$ dimensions of radius proportional to $\sqrt{E}$ and thickness proportional to $1/\sqrt{E}$. Combining the numerical factors into a single constant $C$,
\begin{equation*}
\Omega_\mathrm{gas}(E)= C\, V^n E^{(3n-2)/2}\sim C\, V^n E^{3n/2},
\end{equation*}
where the last step is appropriate when $n\gg 1$.

To motivate our general model for the number-of-states function, we note that the logarithm of $\Omega_\mathrm{gas}(E)$ is, up to logarithmic factors, proportional to the number of particles in the limit of large $n$:
\begin{equation}\label{omegagas}
\log{\Omega_\mathrm{gas}(E)}\sim n\left(\log{V}+\frac{3}{2}\log{E}\right).
\end{equation}
This linear behavior with $n$ is the same as the behavior of $V$ and $E$ in the thermodynamic limit, where the volume and energy per particle are held fixed. The intuition behind the linear behavior of $\log{\Omega}$, for general macroscopic systems, is that these systems can usually be partitioned into identical and nearly independent subsystems of some fixed size. Since $\Omega$ for independent subsystems is multiplicative, it will behave as the power of the number of subsystems. System properties that grow in proportion to the number of degrees of freedom, such as $V$, $E$ and  $\log{\Omega}$, are called \textit{extensive}.

Taking the energy derivative of \eqref{omegagas} we obtain a quantity that behaves, in the thermodynamic limit, as the ratio of two extensive quantities:
\begin{equation}\label{betagas}
\frac{d}{d E}\log{\Omega_\mathrm{gas}(E)}=\frac{3}{2}\left(\frac{n}{E}\right).
\end{equation}
Quantities that are fixed in the thermodynamic limit, such as density and now the quantity above, are called \textit{intensive}. For a general macroscopic system we define the following intensive quantity:
\begin{equation}\label{beta}
\beta(E)=\frac{d}{d E}\log{\Omega(E)}.
\end{equation}
In the case of the weakly interacting gas of particles in three dimensions, $\beta(E)$ is $3/2$ times the inverse mean kinetic energy per particle.

An equivalent and more illuminating restatement of \eqref{beta}, that the energy-derivative of the logarithm of the number-of-states function is intensive, is the following:
\begin{equation}\label{omegafluct}
\Omega(E+E')=\Omega(E)\exp{\left(\beta(E)\,E'\right)}.
\end{equation}
We are justified in keeping just the first two terms of the Taylor series for $\log{\Omega(E)}$ provided the energy fluctuations $E'$ we consider are bounded as we take the thermodynamic limit (since then $E'/E\to 0$). On the other hand, the value of the bound on $E'$ is arbitrary, and so the change in the number-of-states function can be substantial when it exceeds the energy scale defined by $\beta^{-1}$.

\section*{Temperature}

Up to now we have discussed the number-of-states function in mathematical terms, without a physical context. We will arrive at an interpretation of the $\beta$-function by considering two weakly interacting macroscopic systems. By ``weakly interacting" we mean that the Hamiltonian for the joint system is well approximated by the sum $H\approx H_1+H_2$, where the two parts have no variables in common and the neglected terms allow for the exchange of energy between the parts. Two subsystems having this description are said to be in \textit{thermal contact}. By the ergodic hypothesis, the joint phase space distribution of the system, for total energy $E=E_1+E_2$, is
\begin{align*}
\rho&=\rho_0\,\delta(H_1+H_2-E_1-E_2)\\
&=\rho_0\int dE'\;\delta(H_1-E_1-E')\delta(H_2-E_2+E'),
\end{align*}
where $\rho_0$ is the normalization constant. When the integration variable $E'$ is not macroscopic in scale, the joint distribution $\rho$ can be thought of as the product of two distributions, one for system 1 with macroscopic energy $E_1$ and energy fluctuation $+E'$, the other for system 2 with macroscopic energy $E_2$ and energy fluctuation $-E'$. Integrating $\rho$ over phase space and using \eqref{omegafluct}, we obtain
\begin{align*}
1&=\rho_0 \int dE'\; \Omega_1(E_1)\exp{\left(\beta_1(E_1)\,E'\right)}\;\Omega_2(E_2)\exp{\left(-\beta_2(E_2)\,E'\right)}\\
&\propto \int dE' \exp{\left((\beta_1(E_1)-\beta_2(E_2))E'\right)}.
\end{align*}
When $\beta_1-\beta_2>0$, the phase space distribution favors arbitrarily large positive fluctuations $E'$, that is, the transfer of energy from system 2 to system 1. The values $E_1$ and $E_2$ for the two system energies, even in an average sense, are therefore suspect. For such large $E'$ our model for the number-of-states function breaks down, and we should say that the \textit{macroscopic} energy of system 1 has increased and that of system 2 has decreased. The expansions of the number-of-states functions should then be about energies $E_1'>E_1$ and $E_2'<E_2$. To see how this will change the behavior of energy fluctuations, we first need to examine how $\beta(E)$ depends on $E$.

Referring again to the weakly interacting gas model as a guide, we will assume that our general system also has the property exhibited in \eqref{betagas}, that $\beta(E)$ is positive and a decreasing function of $E$. In the scenario above, we would then have $\beta_1(E_1')<\beta_1(E_1)$ and $\beta_2(E_2')>\beta_2(E_2)$, and therefore
\begin{equation*}
\beta_1(E_1)-\beta_2(E_2) > \beta_1(E_1')-\beta_2(E_2').
\end{equation*}
The new difference of $\beta$ functions, if still positive, is smaller and will favor positive energy fluctuations to a lesser extent than the original choice of average system energies. Continuing in this way, we see that there exists a special partitioning of the energy as $E=E_1^*+E_2^*$ such that
\begin{equation}\label{thermequi}
\beta_1(E_1^*)=\beta_2(E_2^*),
\end{equation}
where neither sign of energy fluctuation is favored. Only this partitioning of the total energy establishes average energies for the two subsystems.

We can ask what would happen if the two systems considered above were initially isolated and prepared with energies such that condition \eqref{thermequi} was not satisfied and then brought into thermal contact with each other. The analysis above shows that the joint number-of-states function in that case favors a redistribution of energy such that condition \eqref{thermequi} is restored. What actually happens, in physical terms, is that a macroscopic quantity of energy is transferred from the system with small $\beta$ to the system with large $\beta$. Macroscopic energy transfer without changes in macroscopic parameters, such as volume, is called \textit{heat}. The transfer of heat ceases once the $\beta$ values of the contacting systems are equal.

\textit{Temperature} is operationally defined as the property that two systems must have in common for there to be no transfer of energy (heat). Statistical mechanics defines, quantitatively, the \textit{absolute temperature} $T$ as
\begin{equation*}
k_\mathrm{B}T=1/\beta(E),
\end{equation*}
where Boltzmann's constant $k_\mathrm{B}$ serves to convert the conventional Kelvin (K) units of temperature to units of energy (J):
\begin{equation*}
k_\mathrm{B}=1.3806488 \times 10^{-23}\mathrm{J}/\mathrm{K}.
\end{equation*}
The transfer of heat in the direction of small $\beta$ to large $\beta$ is consistent with everyday experience, of heat flowing from hot to cold (large $T$ to small $T$). Referring to \eqref{omegafluct} we can also give the Boltzmann constant a microscopic interpretation. Suppose we have a macroscopic system at temperature $T=1000$ K; by transferring the microscopic energy $E'=k_\mathrm{B} T\approx 10^{-20}\mathrm{J}$ to the system in the form of heat, its number-of-states is increased by the factor $e=2.718\ldots$ .

\section*{The Boltzmann distribution}

Suppose we have two systems in thermal contact: a truly macroscopic system and a much smaller system. The macroscopic system is so much larger that any energy it exchanges with the smaller system is effective microscopic; its role is simply to establish a temperature $T$. If we are primarily interested in the small system, and the ergodic hypothesis gives us a uniform distribution for both systems in thermal contact, what is the marginal phase space distribution of the small system?
Our model for the number-of-states function of a macroscopic system \eqref{omegafluct} provides the answer to this question.

Let $H(q,\ldots, p,\ldots)$ be the Hamiltonian of the macroscopic system, sometimes referred to as the \textit{thermal reservoir}. This Hamiltonian is weakly coupled to the small system Hamiltonian $H'(q',\ldots,p',\ldots)$. We write the phase space distribution of the weakly coupled systems, with total (macroscopic) energy $E$, as an integral over energy fluctuations $E'$, just as we did above in the discussion of temperature:
\begin{equation*}
\rho(q_1,\ldots,p_1\ldots; q'_1,\ldots,p'_1,\ldots)=\rho_0\int dE'\;\delta(H-E-E')\delta(H'+E').
\end{equation*}
Integrating over just the phase space variables of the macroscopic system and using \eqref{omegafluct}, we obtain the marginal distribution:
\begin{align*}
\rho(q'_1,\ldots,p'_1,\ldots)&=\rho_0\int dE'\;\Omega(E)\exp{\left(\beta(E)\,E'\right)}\,\delta(H'+E')\\
&\propto \exp{\left(-\beta\,H'(q'_1,\ldots,p'_1,\ldots)\right)}.
\end{align*}
The only property of the macroscopic system that has survived is its temperature. This distribution, named after Boltzmann, is far from uniform. Its accuracy, for energy fluctuations potentially spanning many orders of magnitude, is limited only by the degree to which the thermal reservoir has more degrees of freedom. The ``Boltzmann factor", usually written $\exp{(-\Delta E/k_\mathrm{B}T)}$, represents the reduction in the number-of-states function of the reservoir when it gives up energy $\Delta E$ to the smaller system it is in thermal contact with.

\section*{Thermal averages}

For the rest of this lecture we always consider systems in contact with a thermal reservoir. The system Hamiltonian will be called $H$, and its quantum energy levels $E_N(x)$ may depend on an external parameter $x$. We choose to work with quantum states instead of classical Hamiltonian variables because it is through the former that we are able to define the force associated with an adiabatic change in $x$. By writing averages as explicit sums over energy levels we also emphasize the fact that our system can be truly microscopic.

We write the Boltzmann probability of energy state $N$ as
\begin{equation*}
p_N=\exp{(-\beta E_N(x))}/Z,
\end{equation*}
where the normalization factor
\begin{equation*}
Z(\beta,x)=\sum_N \exp{(-\beta E_N(x))}
\end{equation*}
is called the \textit{partition function}. The average of the Hamiltonian, in the thermal or Boltzmann distribution, has a neat expression in terms of $Z$:
\begin{align}\label{aveH}
\langle H\rangle&=\sum_N p_N E_N\\
&=\frac{1}{Z}\sum_N \exp{(-\beta E_N)} E_N\notag\\
&=-\frac{\partial}{\partial \beta}\log{Z}.\notag
\end{align}
Nearly the same kind of expression gives us the thermal average of the force:
\begin{align}\label{force}
\langle F\rangle&=\sum_N p_N \left(-\frac{d E_N}{d x}\right)\\
&=\frac{1}{Z}\sum_N \exp{(-\beta E_N)} \left(-\frac{d E_N}{d x}\right)\notag\\
&=\beta^{-1}\frac{\partial}{\partial x}\log{Z}.\notag
\end{align}

\section*{Entropy}

We can relate the thermal averages for energy and force with the help of another thermodynamic quantity, the \textit{entropy}. Up to the choice of units, the entropy in statistical mechanics is defined exactly as in information theory:
\begin{equation}\label{entropy}
S=k_\mathrm{B}\sum_N p_N\log{(1/p_N)}.
\end{equation}
Multiplying $S$ by the absolute temperature $T$ of the thermal reservoir we obtain something having units of energy. To arrive at an interpretation of this energy, we substitute the Boltzmann probabilities for $p_N$:
\begin{align*}
S&=k_\mathrm{B}\sum_N\left(\frac{\exp{(-\beta E_N)}}{Z}\right)\left(\log{Z}+\beta E_N\right)\\
&=k_\mathrm{B}\log{Z}+\langle H\rangle/T.
\end{align*}
After multiplying by $T$, taking the derivative with respect to the external parameter $x$ and using \eqref{force}, and rearranging, we obtain:
\begin{equation*}
\frac{\partial \langle H\rangle}{\partial x}=T\,\frac{\partial S}{\partial x}-\langle F\rangle.
\end{equation*}
The final step is to integrate this between two values of the external parameter:
\begin{equation*}
\langle H(x_2)\rangle-\langle H(x_1)\rangle=T(S(x_2)-S(x_1))-\int_{x_1}^{x_2}\langle F(x)\rangle dx.
\end{equation*}
In standard thermodynamic notation this takes the form
\begin{equation}\label{deltaU}
\Delta U=T\Delta S+W,
\end{equation}
where $U$ represents the ``internal energy" of the system and $W$ is the work performed \textit{on} the system (by the external mechanism that caused the change in $x$). Because the temperature is held fixed by contact with the thermal reservoir, the process described by \eqref{deltaU} is called \textit{isothermal}.

Whereas the $\Delta U$ and $W$ terms of the thermodynamic relation \eqref{deltaU} are clearly energies, the relationship between energy and the entropy term is more mysterious. In thermodynamics, the product $T\Delta S$ is often written as $Q$, the transfer of energy in the form of \textit{heat}. Statistical mechanics provides a mechanistic basis for heat, and from the definition of entropy \eqref{entropy} we see that the outcome of the transfer of heat energy from the reservoir must be a change in the probabilities $p_N$. Since the Boltzmann probabilities are the marginal distribution of the joint system, to understand these changes we need to consider changes in the reservoir as well.

To help us track the trail of energy in the isothermal process, we note that the premise of ``weak coupling" , between system and reservoir, implies not only that the variables in the two parts are independent, but that the phase-space probability distribution of the joint system is (again approximately) the product of two independent distributions. The entropy of the joint system is therefore the sum of the entropies of the system and reservoir:
\begin{equation*}
S+S_\mathrm{res}=S_0.
\end{equation*}
The joint entropy $S_0$ is constant because the change in $x$ imposed from the outside is adiabatic. From this we conclude $\Delta S=-\Delta S_\mathrm{res}$. Now the reservoir has a uniform distribution over all its states by the ergodic hypothesis, and that implies its entropy is related to the reservoir number-of-states function by
\begin{equation*}
S_\mathrm{res}=k_\mathrm{B}\log{\Omega_\mathrm{res}(E+E')},
\end{equation*}
where, as before, $E$ is the macroscopic energy that establishes the temperature $T$ and $E'$ is the much smaller energy transferred from the system to the reservoir. Using our model \eqref{omegafluct} for the reservoir number-of-states function, we obtain
\begin{equation*}
\Delta S_\mathrm{res}= E'/T,
\end{equation*}
and finally
\begin{equation*}
Q=T\Delta S=-T\Delta S_\mathrm{res}=-E'.
\end{equation*}
We now have a complete accounting of all the energies in the thermodynamic relation \eqref{deltaU}: the change in the system internal energy $U$ is caused both by the input of energy in the form of work $W$ by an external mechanism, and by the flow of energy (heat) $E'$ into the reservoir.

\section*{Thermal elasticity}

It might be a good idea to review the polymer model of Lecture 1 because it will serve as our main example of the concepts just introduced. The energy levels of this model have the large $N$ asymptotic form
\begin{equation*}
E_N(x)\propto (N/V(x))^{2/d},
\end{equation*}
where $d=n-2$ is the number of degrees of freedom for a polymer of $n$ struts and the configuration space volume for polymer extension $x$ is approximately
\begin{equation*}
V(x)\approx V(0)\exp{\left(-\frac{(x/l)^2}{n}\right)}
\end{equation*}
in which we restored the strut length $l$. We will use the proportionality symbol to hide constants that do not depend on parameters such as $x$ and $\beta$.

A polymer in solution is perhaps not as convincing an example of weak coupling to a thermal reservoir as a bottle of gas making thermal contact with its environment only through the walls of the bottle. The merits of our model on a phenomenological level would depend on whether the presence of the solvent introduces configuration-dependent forces, or if this only modifies the parameters, such as the masses, already in the model. Weak coupling would certainly be valid for a polymer in vacuum and interacting only with ambient thermal blackbody radiation. Whether the environment is a solvent or blackbody photons, we will assume that weak coupling applies and the only property of this environment that matters is its temperature.

Our first task is to calculate the partition function $Z$:
\begin{equation}\label{Z}
Z=\int_0^\infty dN\, \exp{(-\beta E_N)}.
\end{equation}
Already we have made two approximations. First, in replacing the sum over $N$ by an integral, we have declared that discreteness of the energy levels is insignificant by our choice of $\beta$ being not too large. More specifically, we limit ourselves to temperatures such that $\Delta E/k_\mathrm{B}T\ll 1$, where $\Delta E$ is a typical level spacing. Second, the upper limit of our integral clearly extends beyond the range of validity of the model. By keeping $\beta$ above some minimum value we can ensure the integral cuts off beyond the point where this is a problem. Our partition function thus will be valid in a range of $\beta$ bounded both above and below.

Using the relation
\begin{equation*}
N\propto V(x) E_N^{d/2},
\end{equation*}
we can change variables in the integral \eqref{Z} :
\begin{align*}
Z(\beta,x)&\propto V(x)\int_0^\infty dE\, E^{d/2-1} \exp{(-\beta E)}\\
&\propto V(x) \beta^{-d/2}.
\end{align*}
The average energy and force then follow from \eqref{aveH} and \eqref{force}:
\begin{align*}
\langle H\rangle&=(d/2) k_\mathrm{B} T\\
\langle F\rangle&=k_\mathrm{B} T\left(\frac{1}{V}\frac{d V}{d x}\right)\approx -\left(\frac{2 k_\mathrm{B} T}{n l^2}\right)x.
\end{align*}

Our result for the average energy depends only on the exponent in Weyl's law for our billiards Hamiltonian. The case of billiards in flat space is also covered by the \textit{equipartition theorem}, which applies to quadratic Hamiltonians and asserts that the average energy is simply $(1/2)k_\mathrm{B}T$ per positive eigenvalue of the quadratic form. A flat space billiard in dimension $d$ has a quadratic kinetic energy with exactly $d$ positive eigenvalues. The fact that the configuration space of our polymer is curved does not change the average energy from the flat space value.

The average force generated by the polymer and acting on its moveable end has exactly the form of a Hookean spring, with stiffness proportional to the absolute temperature. A hot polymer makes a stiffer spring because there is a steeper entropic penalty for being extended when the temperature is high. We can see this also by evaluating the entropy difference,
\begin{equation*}
T\Delta S=T(S(x)-S(0))=-\left(\frac{k_\mathrm{B} T}{n l^2}\right)x^2
\end{equation*}
and noting that $\Delta U=0$ (since $\langle H\rangle$ is independent of $x$) in \eqref{deltaU}:
\begin{equation*}
0=-\left(\frac{k_\mathrm{B} T}{n l^2}\right)x^2+W.
\end{equation*}
Thus all the work performed by the external mechanism on extending the polymer is cancelled by the loss in entropy.

\section*{Free energy}

Statistical mechanics makes yet another contact with thermodynamics through the \textit{Helmholtz free energy}, defined as
\begin{equation*}
F=-\beta^{-1}\log{Z}=\langle H\rangle -T S.
\end{equation*}
Both $\langle H\rangle$ and $T S$ (through its change $Q=T \Delta S$) have macroscopic, thermodynamic interpretations. In statistical mechanics these quantities are defined microscopically through the probabilities of the individual energy levels, $p_N$. The connection between the two points of view takes the form of a minimum principle.

Suppose we did not know that the probabilities of a system in contact with a thermal reservoir have the Boltzmann form. Instead we propose that the probabilities are determined by the property that they optimize something. Maximizing the entropy gives the uniform distribution, while minimizing the energy just gives the lowest energy state(s). As something in between, we try minimizing the free energy $F$ with respect to the probabilities. Because the probabilities sum to 1, this is a constrained minimization problem that we solve by the method of Lagrange multipliers. Thus we minimize the function
\begin{equation*}
F-\lambda\sum_N p_N=\sum_N p_N(E_N+\beta^{-1}\log{p_N}-\lambda)
\end{equation*}
with respect to the $p_N$ subject to no constraint and general Lagrange multiplier $\lambda$, and then solve for $\lambda$ such that the $p_N$ sum to 1.
The result of this easy exercise is that the $p_N$ have exactly the Boltzmann form.

The appeal of optimization principles is that they provide a basis for intuition. In the case of the free energy minimization principle we see how temperature tips the scale in favor of energy or entropy. At low temperatures the principle emphasizes energy, and this is what a system will minimize. When the temperature is high, entropy gains the upper hand.

\section*{Thermal equilibrium}

Although statistical mechanics is the study of mechanics from which time has been eliminated, we should not overlook the significance of time scales. In the previous Lecture we saw that a process is only adiabatic when it is carried out on a time scale that is long on the scale of the system's dynamics. Another time scale is relevant when our system is in contact with a thermal reservoir. This scale is set by the rate that energy can flow between system and reservoir. As we get ever closer to the ideal of weak coupling, this rate of thermal equilibration goes to zero and the time scale diverges.

In this Lecture we found that the force generated by a freely-jointed polymer is changed when it is in contact with a thermal reservoir, over what it was in isolation. In isolation the polymer energy increases when extended, while in a solvent at temperature $T$ its energy stays constant at a value set by $T$. When the coupling to the solvent is weak, it may be possible to extend the polymer quickly and not maintain thermal equilibrium with the solvent. In that case the polymer will gain energy, as in the adiabatic process. Processes are often described as adiabatic for just this reason: there is insufficient time to transfer energy to and from the thermal environment. 

\section*{Problems for study}

\subsection{Naturalness of logarithms}

Does physics ``know" about the number $e$, the base of the natural logarithms? Information theory, by convention, defines entropy with the base 2 logarithm. Is the entropy of statistical physics more natural, as an outgrowth of natural law? Not unrelated is the Boltzmann distribution, in which $e$ also figures prominently.

\subsection{Ideal gas as multi-dimensional billiards}
An ideal gas of $n$ identical point masses in a volume $V$ is equivalent to a billiard (1 particle) in $3d$ dimensions and volume $V^n$. Use this equivalence and Weyl's asymptotic law for the $N$th energy level to calculate the number-of-states function $\Omega_\mathrm{gas}(E)$.

As another application of Weyl's law, show that the energy of the gas satisfies
\begin{equation*}
E(V)=E(V_0)\left(\frac{V_0}{V}\right)^{2/3},
\end{equation*}
when the volume of the gas is changed adiabatically ($V_0$ is an arbitrary reference volume).

\subsection{Slightly non-ideal gas}
The \textit{hard-sphere gas} model is a multi-dimensional billiard, just like the ideal gas, the only difference being that the configuration space is restricted by the constraint that all particles have separation at least $2b$, where $b$ is the hard-sphere radius. The volume of the billiard is therefore not equal to $V^n$, but something smaller. Find an approximation for the reduced billiard volume that applies when the total volume covered by spheres is much smaller than $V$.

\subsection{Ideal gas in contact with thermal reservoir}
Revisit the ideal gas system, but now in contact with a thermal reservoir at temperature $T$. Use the energy levels from the second problem to calculate the partition function, and then \eqref{aveH} to find the average energy. Is your answer consistent with the equipartition theorem?

The analog of the external parameter $x$, by which we defined force in the polymer model, is the volume $V$ occupied by the gas. When the gas is in the energy state $E_N$ its pressure is
\begin{equation*}
p=-\frac{\partial E_N}{\partial V}.
\end{equation*}
Find the thermal average $\langle p\rangle$ in analogy with the calculation of average force for the polymer. Does the result surprise you?

Compare the formula for the pressure of the gas at temperature $T$ with the formula for the gas in isolation, that is, when the gas is in a particular energy state. Select the energy state so the two gases have the same pressure $p_0$ at the reference volume $V_0$. 

\subsection{Isothermal compression of the ideal gas}
An ideal gas in contact with a thermal reservoir at temperature $T$ is compressed from volume $V_1$ to volume $V_2$. Calculate the change in internal energy, the change in entropy, and the work performed on the gas using
\begin{equation*}
W=-\int_{V_1}^{V_2} \langle p\rangle dV.
\end{equation*}
Check that the general thermodynamic relation \eqref{deltaU} holds.

\subsection{Cosmic microwave background}
The universe is said to be filled with microwave radiation at absolute temperature $T=2.73$ K. How is it possible to say a system as large as the universe is at some temperature when nothing bigger exists that could be performing the role of thermal reservoir? Also, the temperature of this radiation is claimed to have been much higher in the distant past --- how is that possible?

\lecture{Macroscopic order}

\section*{Macroscopic manifestations of microscopic order}

In addition to providing a microscopic foundation for thermodynamics, statistical mechanics is also instrumental in showing how physics on the micro-scale is translated into macroscopic phenomena. The classic example of elementary interactions transcending many orders of magnitude in scale is the origin of crystal facets, and the related fact that crystals scatter radiation much like a macroscopic mirror but according to rules derived directly from long range geometric order in the atomic structure. We will examine this example by way of a simplified model. Statistical mechanics serves us in two ways: to explain the mechanism of microscopic order, and also to set limits on the degree to which microscopic order translates to order on the macro-scale.

\section*{Hard spheres: microscopic order}

In materials such as silicon, the origin of crystallinity is directly linked to the bonding geometry of the constituent atoms. Atomic order in these materials comes about through the minimization of energy. In its crystalline form, the energy of silicon is about 5 eV per atom lower than it is when the atoms form a gas. Silicon atoms in contact with a reservoir whose thermal energy scale $k_\mathrm{B} T$ is several times smaller than 5 eV (say a reservoir at 1,000 K) will have such a high probability of being in the unique crystalline configuration that there is no further role for statistical mechanics.

Order, even on a microscopic scale, can happen through entropy as well. The most studied example of this phenomenon, called \textit{order by disorder}, is the hard sphere system. In this model there is no energy scale: the energy of any configuration of nonintersecting spheres is the same, which for convenience we take to be zero. The noble gas atoms (helium, neon, etc.) are well modeled by this system, because the pair potential energy rises sharply below a certain distance; at low temperatures such configurations are simply excluded as they would be, for any temperature, in the hard sphere system.

We will show the results of some numerical experiments on the hard disk system in two dimensions to explain the mechanism of order by disorder. Figure 1 shows the initial positions of 63 hard disks in a square box. They are separated by small gaps from each other and the walls of the box. After giving them random velocities we run the equations of motion and see what happens. Packed as they are in a tight square lattice, one might expect the disks to ceaselessly rattle around, maintaining their average positions. One of the disks has been removed to check whether this hypothesis is robust. If the square crystal maintains its integrity, then the missing disk, or ``vacancy", should diffuse around without disturbing the crystal as a whole. But as Figure 2 shows, this is not what happens.

\begin{figure}
\includegraphics{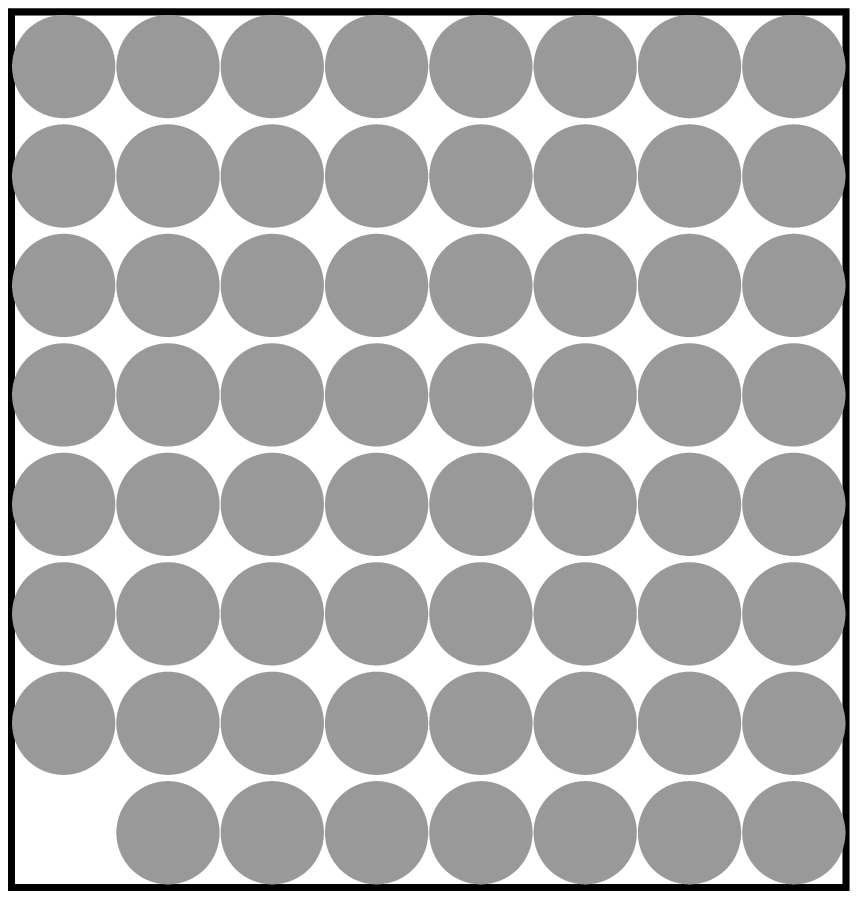}
\caption{Initial positions of 63 hard disks. Upon being given random velocities the disks rearrange themselves over time into more loosely packed configurations, such as the one shown in Figure 2.}
\label{}
\end{figure}

After a relatively short time, measured by numbers of collisions with neighboring disks, the square crystal structure settles into a more loosely packed structure, such as shown in Figure 2. We can interpret the disintegration of the square lattice over time as a manifestation of free energy minimization with respect to the static probabilities of statistical mechanics. In the case of hard spheres (disks), free energy is minimized, at any temperature, by maximizing the entropy. When the disks are packed as a square lattice with small gaps, there is very little ``free volume" of movement for each disk: the entropy is very small. By adopting a different kind of order, and favoring configurations that have a greater capacity for disorder, the entropy is higher and the free energy is lower.

\begin{figure}
\includegraphics{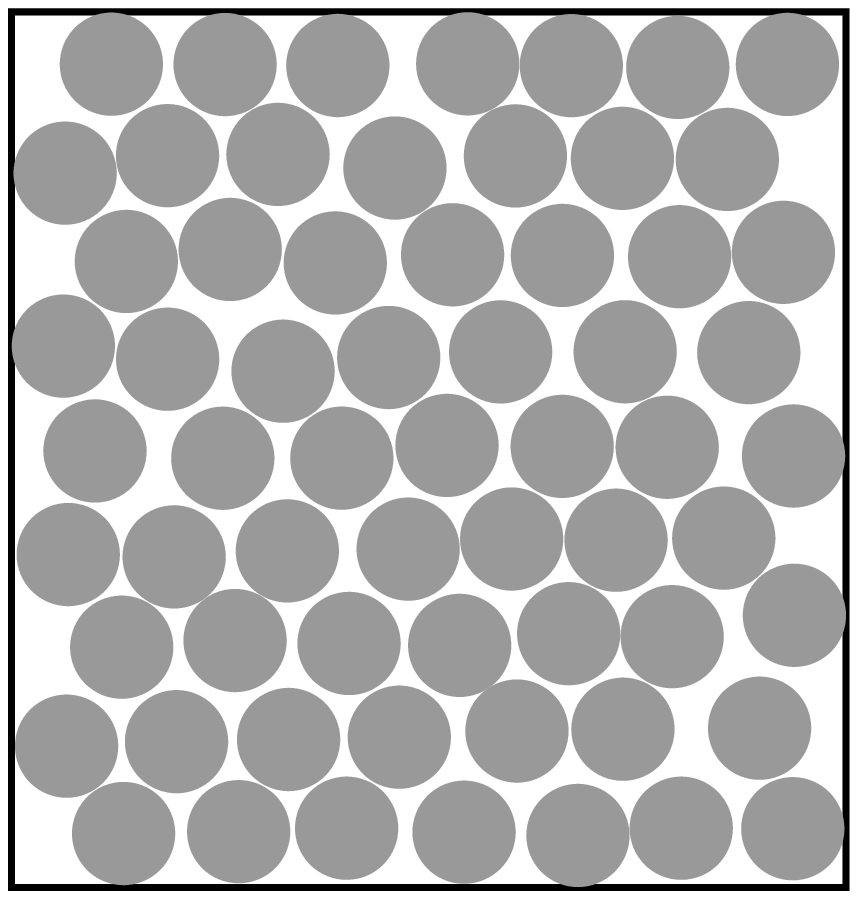}
\caption{A typical configuration of 63 hard disks has nine rows of seven disks in a hexagonal arrangement.}
\label{}
\end{figure}

Although the disks in Figure 2 appear disordered, they too possess crystalline order. A hexagonal lattice is evident when we aggregate positions over time, as shown in Figure 3. That the hexagonal crystal must prevail above some value of the density of disks is reasonable, because exactly at the maximum packing density the hexagonal arrangement is the \textit{only} allowed configuration of disks. The question for statistical mechanics to answer is whether this hexagonal order persists, even at densities below this maximum density of ``close-packed" disks.

\begin{figure}
\includegraphics{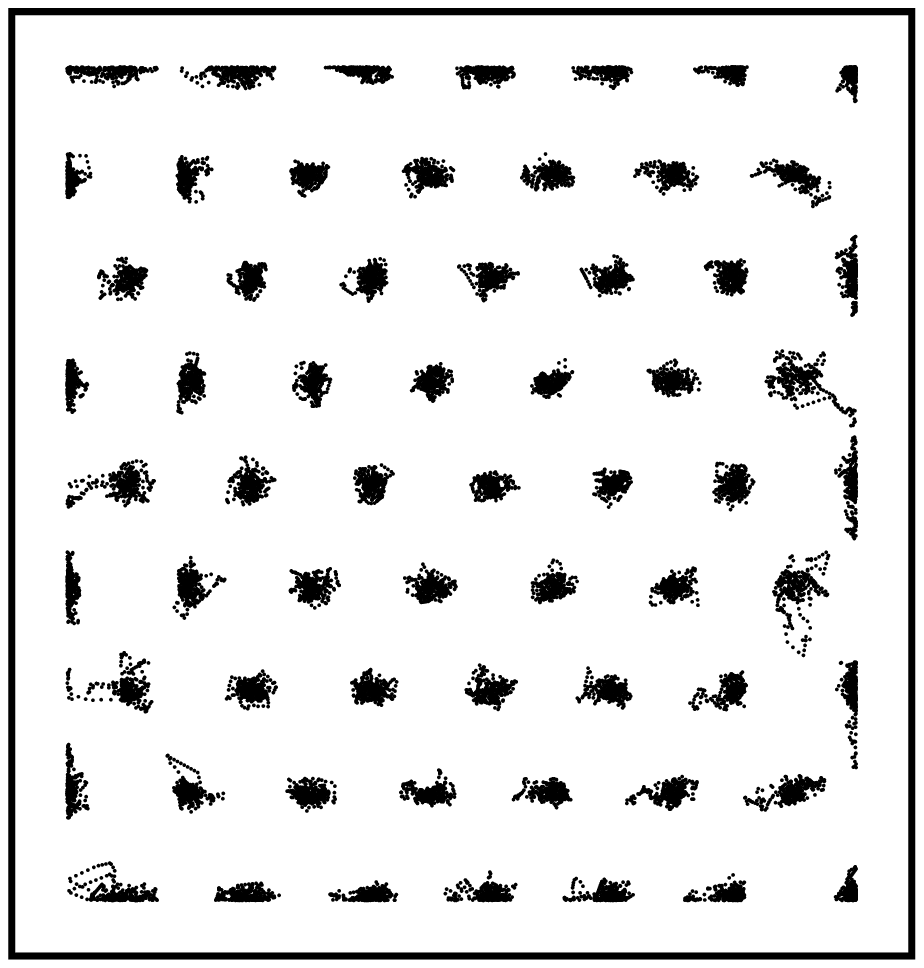}
\caption{Scatter plot, over a period of many collision times, of the distribution of disk centers. The averages of these distributions form a hexagonal lattice.}
\label{}
\end{figure}

A simple plausibility argument, for the existence of crystalline order below the close-packing density, goes like this. Let $v=V/n$ be the volume per disk for a system of $n$ disks, where ``volume" in this case the total area of the system. At close packing a system of unit-radius disks has $v=v_c=2\sqrt{3}$. What could the behavior of the total configuration space volume $V_n$ of a system of $n$ disks look like, as $v$ tends to $v_c$? This quantity is a function of the parameter $v$, vanishes when $v=v_c$, and scales as volume in $2 n$ dimensions. A reasonable candidate (for $n$ large so boundary effects are small) is
\begin{equation}\label{asymvol}
V_n\sim \left(c\,(v-v_c)\right)^n,
\end{equation}
the only unknown being the constant $c$. We can even estimate $c$ by interpreting $c\,(v-v_c)$ as the free volume available to one disk when its neighbors are fixed at their average positions. This gives $c= 1$. For a better estimate of $c$ we would have to consider correlations in the motions of the disks. But whatever the true value of $c$, the fact that the asymptotic form \eqref{asymvol} makes reference to a specific crystalline configuration (and applies for $v>v_c$) is consistent with the proposition that there is order even below the close-packed density.

Statistical mechanics plays a greater role in determining the crystalline structure of hard spheres in three dimensions because geometry by itself is inconclusive. In three dimensions there is not a unique densest packing but an infinite set of packings differing in the sequence that three kinds of hexagonally ordered layers are stacked. As shown in Figure 4, above any one layer the next layer must be of one of the other two types. Referring to the three kinds of layer as $A$, $B$ and $C$, there are two stacking sequences that prevail in actual crystals: the fcc or \textit{face-centered-cubic} sequence $ABCABC\ldots$ and the hcp or \textit{hexagonally-close-packed} sequence $ABAB\ldots$ . It is widely believed that the asymptotic form \eqref{asymvol} holds for hard spheres \cite{Stillinger}, and it is only the constant $c$ that distinguishes the different stacking sequence structures. Extensive numerical calculations were necessary to resolve differences among these constants and it is now believed that the fcc and hcp sequences give the extremes in the spectrum of values. The configuration space volumes for these extremes are still very close \cite{Elser1}:
\begin{equation*}
c_\mathrm{fcc}/c_\mathrm{hcp}\approx 1.00116.
\end{equation*}
Since the entropy of the hard sphere system is just $k_\mathrm{B}$ times the logarithm of $V_n$ (plus a term that just depends on the temperature), the two sphere packings have free energy difference
\begin{equation*}
(F_\mathrm{fcc}-F_\mathrm{hcp})/n= -k_\mathrm{B}\, T\log{\left(\frac{c_\mathrm{fcc}}{c_\mathrm{hcp}}\right)}\approx -0.00116\, k_\mathrm{B} T.
\end{equation*}
We can compare this purely entropic contribution to the free energy difference to any energetic contributions we neglected when modeling our noble gas atoms as hard spheres. Given the small size of the entropic contribution, the selection between the fcc and hcp crystal forms is most likely determined by energy.

\begin{figure}
\includegraphics{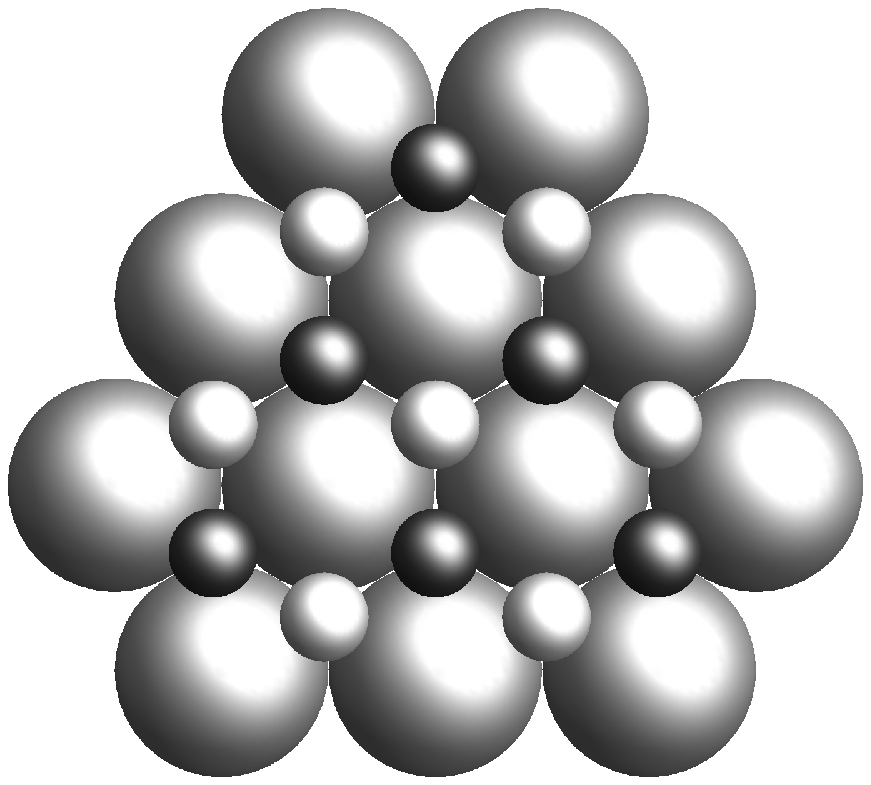}
\caption{Layer stacking in close-packed spheres. The large spheres show a single hexagonal layer, of type $A$. The layer above this layer is again hexagonal, but there is a choice of centering the spheres on the small light colored spheres, type $B$, or the small dark spheres, type $C$.}
\label{}
\end{figure}

A more difficult question is determining the largest volume per disk (or sphere) below which crystalline order first appears. This is a much studied problem and well beyond the scope of these lecture. We can at least develop an intuitive understanding of such order-disorder \textit{phase transitions} by once again considering free volumes. In the disordered or \textit{gas phase}, the disks are only weakly correlated. Even so, a typical disk will be surrounded by a number of nearby disks that strongly limit the size of its free volume. The entropy of these disordered configurations will be diminished as a result of these ``caging" effects. To partially mitigate the entropic penalty of caging, disks may coordinate on a set of mutually beneficial average-positions. While this introduces an entropic penalty, by constraining the average positions to a crystal, the gain in free volumes enabled by these average positions can result in a net entropy gain. The density where this social contract among disks first takes effect marks the phase transition to the ordered, or \textit{solid phase}.

\section*{Hard spheres: macroscopic order}

In our argument for microscopic crystalline order in the hard sphere system we considered the limit $v\to v_c$ as $n$ was held fixed. When addressing macroscopic properties we are actually interested in the opposite limit: $v$ fixed and $n\to\infty$. Does crystallinity survive in the thermodynamic limit?

Numerical experiments with disks casts doubts on the existence of macroscopic order. Even when the gaps between disks are small in the perfectly ordered arrangement, there can be large net motions of disks when  gaps are slightly compressed or expanded over large regions. In large systems we often observe that rows of disks deviate from straight lines by several disk diameters, sometimes opening gaps wherein the hexagonal microscopic order is lost.

To address the large scale motions of disks and their threat to macroscopic order, we construct a model for just these degrees of freedom. We start with a small system, such as the one in Figure 2, whose crystal structure is under our control by means of boundary conditions. Let $\mathbf{r}$ denote the average positions in the crystal with perfect hexagonal symmetry and volume $V$. We can displace the average positions by a small linear transformation
\begin{equation*}
\mathbf{r}'=\mathbf{r}+\mathbf{u}(\mathbf{r}) = \mathbf{r}+\mathbf{A}\cdot\mathbf{r}
\end{equation*}
by applying the same displacement to the boundary of our system. From the work performed by the boundary on the disks we can determine, in principle, the free energy change associated with any linear distortion of the crystal structure.

The four degrees of freedom of the matrix $\mathbf{A}$ decompose into expansion, shear and rotation modes:
\begin{equation*}
\mathbf{A}=\left[
\begin{array}{cc}
a_0+a_1&a_2+a_3\\
a_2-a_3&a_0-a_1
\end{array}
\right].
\end{equation*}
The parameter $a_0$ corresponds to uniform expansion (or compression for negative $a_0$), $a_1$ and $a_2$ are the two orthogonal modes of shear, and $a_3$ generates a rotation. By general principles we can argue that the expansion of the free energy to second order takes the form
\begin{equation}\label{F(A)}
F(\mathbf{A})/V-F(0)/V=-2 p a_0+\kappa_1 a_0^2+\kappa_2 (a_1^2+a_2^2) +\cdots .
\end{equation}
The rotation parameter $a_3$ does not appear at all because the free energy is unchanged when the system is rotated. Only the expansion parameter may appear to first order because a linear term in $a_1$ or $a_2$ is inconsistent with hexagonal symmetry having the lowest free energy (at constant volume). The coefficient of the $a_0$ term, proportional to the volume derivative of the free energy, is just the pressure (its analog in two dimensions).  Finally, the two shear mode stiffnesses are equal by the assumed hexagonal symmetry of the crystal.

To model the macroscopic system we interpret the free energy change \eqref{F(A)} of the small system with constant distortion $\mathbf{A}$ as the local change in the free energy density of a large system with variable $\mathbf{A}$. Since
\begin{align*}
a_0&=(\partial_x u_x+\partial_y u_y)/2\\
a_1&=(\partial_x u_x-\partial_y u_y)/2\\
a_2&=(\partial_x u_y+\partial_y u_x)/2,
\end{align*}
the free energy of the large system takes the following form as a functional of the displacement field $\mathbf{u}(\mathbf{r})$:
\begin{equation}\label{continuum}
F(\mathbf{u})=\int d^D\mathbf{r} \left( -p\,\nabla\cdot\mathbf{u} +C_{ijkl}(\partial_i u_j)(\partial_k u_l) \right)
\end{equation}
Although we derived this for the case of disks in dimension $D=2$, in other dimensions (and different packing geometries) only the details of the  elasticity tensor $C$ are changed.

We can think of the integrated free energy density \eqref{continuum} as the Hamiltonian of a new system whose variables are the displacement field $\mathbf{u}(\mathbf{r})$. This Hamiltonian has no momentum variables, and it is not possible or even correct to try to derive equations of motion for the displacement field. The correct interpretation of \eqref{continuum} is that it provides an efficient computation of the Boltzmann distribution for systems so large that the displacement field has become a complete specification of the system's degrees of freedom, the microscopic details having been absorbed by the definitions of the pressure and the elasticity tensor. Since the field $\mathbf{u}(\mathbf{r})$ vanishes on the boundary of the macroscopic system, the pressure term in \eqref{continuum} integrates to zero. The squared-gradient form that remains appears in many contexts, not just hard spheres, and is our focus in the next section.

\section*{Order in the height model}

The \textit{height model} is a model that captures the key elements of long range order in a variety of systems. When applied to macroscopic order in the hard sphere system, the ``height" corresponds to a single component of the elastic displacement field. In the next section we will see that it can also represent something much more abstract.

The height model has only one interesting parameter: the dimension of space $D$. When $D=2$ the variables of the height model, $h(\mathbf{r})$, might represent the actual height of a surface above points $\mathbf{r}$ in the plane. The case $D=1$ models a random walk, when the single space coordinate is reinterpreted as time and $h(t)$ represents the position of the walker at time $t$. In fact, we can think of the height model in $D$ dimensions as a generalization of the random walk, where $D$ is not the dimension of the space that is being walked within, but the dimensions of the entity that is ``walking".

We construct new variables for the height model from the amplitudes $h_\mathbf{k}$ of plane-wave modes:
\begin{equation}\label{modes}
h(\mathbf{r})=\sum_\mathbf{k} e^{i\mathbf{k}\cdot\mathbf{r}} h_\mathbf{k}
\end{equation}
Because the height is real-valued, the complex amplitudes satisfy the constraint $h_{-\mathbf{k}}=h_\mathbf{k}^*$. The mode wave-vectors $\mathbf{k}$ are chosen with each component an integer multiple of $2\pi/L$ so the height is a periodic function on a hyper-cubic domain of volume $V=L^D$. We exclude the $\mathbf{k}=0$ mode because it does not contribute to the free energy of the height model. Since the model is only meant to address macroscopic properties, we place an upper cutoff $k_\mathrm{max}=k_0$ on the magnitudes of the wave-vectors. The exclusion of  $\mathbf{k}=0$ means that there is effectively also a lower cutoff, $k_\mathrm{min}=2\pi/L$. We are not so much concerned with the precise values of these cutoffs, but the more salient fact that only the lower cutoff scales in a certain way with the system size $L$.

In our analysis of the height model we will encounter sums over all the wave-vectors, a finite set, of functions that are insensitive to the discreteness of this set. We can therefore approximate such sums by integrals:
\begin{equation*}
\sum_\mathbf{k}\;\cdots\;\approx \int_{k_\mathrm{min}<\|\mathbf{k}\|<k_\mathrm{max}} \rho\, d^D\mathbf{k}\;\cdots,
\end{equation*}
where $\rho=V/(2\pi)^D$ is the density of modes.

The Hamiltonian of the height model is the integral of the free energy density of some microscopic model. As in the case of the entropic elasticity of hard spheres, we adopt the squared-gradient form:
\begin{align*}
H&= \frac{\kappa}{2}\int_V d^D\mathbf{r}  \;\|\nabla h\|^2\\
&= \frac{\kappa}{2}\sum_\mathbf{k} \; V\,\|\mathbf{k}\|^2\, |h_\mathbf{k}|^2.
\end{align*}
We should think of this as the first term in an expansion, the higher derivative terms being smaller in the large-scale, long-wavelength regime we are interested in. The free energy density is limited to just this one term and a single stiffness parameter $\kappa$, because we are assuming rotational symmetry.

With the variables and Hamiltonian of the model defined, we are in a position to calculate various properties of interest. We will focus on a single property, the probability distribution of the height above a particular point $\mathbf{r}_0$:
\begin{equation*}
p(h_0)=\langle \delta(h(\mathbf{r}_0)-h_0)\rangle.
\end{equation*}
The angle brackets denote the average with respect to the Boltzmann distribution for the height model Hamiltonian $H$. Using the Fourier representation of the delta function, we can express the probability distribution as
\begin{equation*}
p(h_0)=\frac{1}{Z(0)}\int_{-\infty}^{+\infty}\frac{dq}{2\pi} e^{-i q h_0}\,Z(q),
\end{equation*}
where
\begin{equation*}
Z(q)=\prod_{\mathbf{k}^+}\left( \int_\mathbb{C}d h_\mathbf{k} \exp{\left(2 i q\, \mbox{Re}(e^{i\mathbf{k}\cdot\mathbf{r}_0}h_\mathbf{k})-  \beta\kappa V\,\|\mathbf{k}\|^2\, |h_\mathbf{k}|^2 \right)}\right),
\end{equation*}
and the product is over just one representative of each $(\mathbf{k},-\mathbf{k})$ pair. By rotating the phase of the complex integration variable $h_\mathbf{k}$ we see that the integral does not depend on the position $\mathbf{r}_0$, in agreement with the translational invariance of the probability distribution. Upon performing the Gaussian integrals for each $\mathbf{k}^+$ we obtain
\begin{equation*}
Z(q)=\prod_{\mathbf{k}^+}\left(\frac{\pi}{\beta\kappa V \|\mathbf{k}\|^2}\right)\exp{\left(-\frac{q^2}{\beta\kappa V \|\mathbf{k}\|^2}\right)},
\end{equation*}
and from that the Fourier representation of the probability:
\begin{equation*}
p(h_0)=\int_{-\infty}^{+\infty}\frac{dq}{2\pi} e^{-i q h_0}\exp{\left(-\frac{1}{2}\sum_{\mathbf{k}}\frac{q^2}{\beta\kappa V \|\mathbf{k}\|^2}\right)}.
\end{equation*}
The final form of the distribution is Gaussian
\begin{equation*}
 p(h_0)=\frac{1}{\sqrt{2\pi\sigma^2}}\exp{\left(-\frac{h_0^2}{2\sigma^2}\right)},
\end{equation*}
where the width $\sigma$ takes three asymptotic forms, depending on dimension, when the sum over modes,
\begin{equation*}
\sum_{\mathbf{k}}\frac{1}{\|\mathbf{k}\|^2}\approx \frac{V}{(2\pi)^D}\int_{k_\mathrm{min}}^{k_\mathrm{max}}\Omega_D\, k^{D-3} dk,
\end{equation*}
is evaluated in the asymptotic limit $k_\mathrm{min}\ll k_\mathrm{max}$ :
\begin{equation}\label{heightsigma}
\sigma^2\propto\left\{
\begin{array}{ll}
L/(\beta \kappa)\,,& D=1\\
\log{(k_0 L)}/(\beta \kappa)\,,& D=2\\
k_0^{D-2}/(\beta \kappa)\,,& D>2.
\end{array}\right.
\end{equation}
The proportionality hides numerical factors ($2\pi$, spherical surface areas $\Omega_D$) to emphasize the dependence on the system size $L$.

Since $D=1$ corresponds to the ordinary random walk, it comes as no surprise that we recover the well known $\sigma\propto\sqrt{L}$ growth of the root-mean-square ``height" with the ``time" $L$ of the walk. The important lesson here is that above $D=2$ the ``order" in the height is perfect in the sense that the width $\sigma$ is bounded --- stays microscopic --- as we take the limit $L\to\infty$.

The ordered, system-wide value of the height in our model was arbitrarily set to zero in our mode expansion
\eqref{modes} when we omitted the $\mathbf{k}=0$ mode. Because the Hamiltonian does not depend on this mode, height fluctuations that do not grow with $L$ would have been obtained for any value of the ordered height in dimension $D>2$. This fact makes the height model one of the simplest examples of \textit{spontaneous symmetry breaking}. The symmetry being broken is the uniform translation of the height over the entire system. Above any point, the distribution of heights has a fixed distribution in the limit of infinite system size. The mean of these heights can be determined even from a single configuration of the surface $h(\mathbf{r})$, since, as further analysis shows, the fluctuations at different points are only weakly correlated. As a result, the value of the height spontaneously selected by the system can be determined with arbitrary precision, and from a single observation, as the system size increases.

Taking the height variable to be a component of the elastic displacement field $\mathbf{u}(\mathbf{r})$ for hard disks or spheres, we see that true long range order exists only in the case of spheres in $D=3$. For disks in $D=2$ the amplitudes of thermal displacement fluctuations grow logarithmically with system size and therefore the degree of order is borderline.

\section*{Random tilings}

The height model makes a surpassing appearance in discrete models of solid ordering. Although the main application is to \textit{quasicrystal order} in two and three dimensions, the general idea can be explained with a simple model in one dimension.

\begin{figure}
\includegraphics{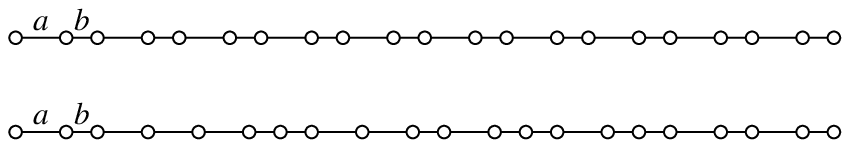}
\caption{Periodic (top) and random (bottom) sequences of two lengths $a$ and $b$.}
\label{}
\end{figure}

Suppose we have a solid in one dimension comprised two two rigid motifs (\textit{e.g.} molecules) of length $a$ and $b$. When these alternate in the structure, as shown in Figure 5, the resulting structure has period $a+b$. Diffraction experiments provide a direct probe of this order. For simplicity, suppose the motifs are bonds of two lengths between identical atoms and that these atoms scatter radiation. Let $X$ be the set of atom positions, and we are interested in experiments where the size $N$ of this set is very large. The \textit{structure factor}, defined by
\begin{equation*}
S(q)=\sum_{x\in X}\exp{(i q x)}
\end{equation*}
gives the amplitude of radiation scattered by the entire system. Here $q$ is the momentum change of the radiation in the scattering process, determined both by the wavelength and the scattering angle\footnote{In $D=1$ there are only two ``angles", corresponding to the sign of $q$.}. When $q$ is an integer multiple of $2\pi/(a+b)$
the terms in the sum repeat and $S(q)$ is proportional to $N$, the number of terms in the sum. This scaling of the structure factor --- the phenomenon of \textit{Bragg peaks} at special $q$'s --- is the signature of periodic order.

Figure 5 also shows a more random sequence of the two motifs. What can we say about the behavior of the structure factor for these? We will find, that with relatively weak assumptions about the degree of disorder, even these structures exhibit Bragg peaks.

A systematic way to analyze the diffraction properties of general sequences of the two lengths is to embed the structure in two dimensions, as shown in Figure 6. Let $a=\cos{\alpha}$ and $b=\sin{\alpha}$ be the two lengths. We are interested in the case where $a$ and $b$ are incommensurate; if they had a common multiple $c$, then we would trivially get a Bragg peak for every $q$ that is a multiple of $2\pi/c$. The construction shown in Figure 6 is to project edges of the square graph onto a line making angle $\alpha$ with the horizontal axis. Any sequence of lengths $a$ and $b$ thus corresponds to a path through the square lattice. The line onto which we project may only pass through a single lattice point, the origin, since otherwise $a$ and $b$ would be commensurate. Similarly, from any projected lattice point on the line we can reconstruct a unique lattice point.

\begin{figure}
\includegraphics{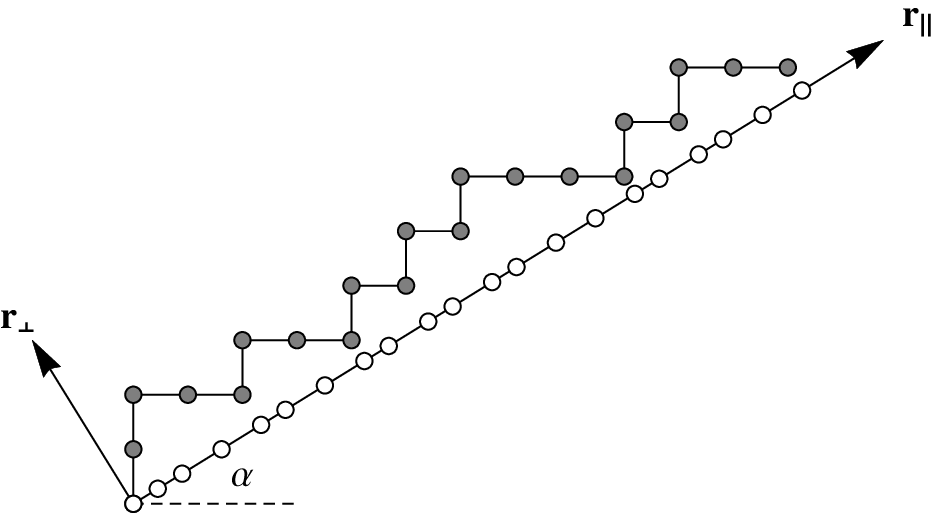}
\caption{Projection of a lattice path onto the line $\mathbf{r}_{||}$, forming the sequence of lengths $\sin{\alpha}$ and $\cos{\alpha}$.}
\label{}
\end{figure}

Let $\mathbf{r}$ be a general point of the square lattice. We can express $\mathbf{r}$ as the sum of its projection on the line, $\mathbf{r}_{||}$, and its orthogonal complement, $\mathbf{r}_\perp$. Also, let $\mathbf{q}$ be a general point of the lattice dual to the square lattice, another square lattice, and scaled by $2\pi$. Thus for any $\mathbf{r}$ and $\mathbf{q}$ we have
\begin{equation}\label{duallattice}
\exp{(i\mathbf{q}\cdot\mathbf{r})}=1.
\end{equation}
We can decompose any $\mathbf{q}$ into the same pair of orthogonal spaces as $\mathbf{r}$ and from \eqref{duallattice} infer the relationship
\begin{equation*}
\exp{(i\mathbf{q}_{||}\cdot\mathbf{r}_{||})}=\exp{(-i\mathbf{q}_{\perp}\cdot\mathbf{r}_{\perp})}.
\end{equation*}

Using the square lattice construction we can identify Bragg peaks of the original two-length sequence in one dimension and calculate their strength. Let $\mathbf{R}$ be a particular set of lattice points forming a path on the lattice and whose projections, the set $\mathbf{R}_{||}$, correspond to the atom positions in the diffraction experiment. By the incommensurate property of the projection, there is a bijection between elements of $\mathbf{R}_{||}$ and the elements of $\mathbf{R}_{\perp}$, the projection of the lattice path onto the orthogonal space. Let $\mathbf{q}_{||}$ be the projection of some dual lattice vector and consider the structure factor with this as the momentum change in the diffraction experiment:
\begin{align}\label{quasistrucfac}
S(\mathbf{q}_{||})&=\sum_{\mathbf{r}_{||}\in\mathbf{R}_{||}} \exp{(i\mathbf{q}_{||}\cdot\mathbf{r}_{||})}\notag\\
&=\sum_{\mathbf{r}_{\perp}\in\mathbf{R}_{\perp}} \exp{(-i\mathbf{q}_{\perp}\cdot\mathbf{r}_{\perp})}.
\end{align}
From the second line we can see how the terms in the sum can be made to combine to give a structure factor that grows as $N$, the number of terms in the sum. Most directly, we can constrain the lattice paths to always lie in a strip of finite extent in $\mathbf{r}_{\perp}$ and consider dual lattice vectors for which the $\mathbf{q}_{\perp}$ is small. Such dual lattice vectors always exists, because the projection subspaces are incommensurate. With these bounds on $\mathbf{r}_{\perp}$ and $\mathbf{q}_{\perp}$ in place, we see that the structure factor does indeed grow as $N$, and there is a Bragg peak at $\mathbf{q}_{||}$ in the diffraction experiment.

Constraining the lattice path to be bounded in $\mathbf{r}_{\perp}$ seems unphysical, since the atoms arranged in $\mathbf{r}_{||}$ do not have access to the geometrical construction that reveals their $\mathbf{r}_{\perp}$. Is there a statistical mechanism that achieves the same thing?

Consider a long lattice path comprised of $N_a$ horizontal and $N_b$ vertical edges. The average slope of the path will match that of the $\mathbf{r}_{||}$ subspace when $N_a/N_b\approx \tan\alpha$. Assuming the two motifs in our 1-dimensional solid have this relative concentration, a simple model might be that all their arrangements are energetically so similar on the thermal energy scale $k_\mathrm{B} T$ that they occur with equal probability. This still leaves open the possibility of large-scale concentration fluctuations within the material, whose effect on the diffraction we turn to next.

The method of analysis follows closely our analysis of long range order in the hard sphere system. The macroscopic region occupied by the solid is partitioned into microscopic domains characterized by local variations in the slope of the lattice path. In each domain we calculate the free energy and its dependence on the local slope. This then becomes the free energy density of a macroscopic model that is integrated to give the probability of arbitrary paths, now described by smooth curves $\mathbf{r}_\perp(\mathbf{r}_{||})$.

Consider a microscopic domain of size $\Delta\mathbf{r}_{||}$ in which the $\mathbf{r}_{\perp}$ projection changes by $\Delta\mathbf{r}_{\perp}$. By geometry, these are related to the number of edges in the path that are horizontal, $n_a$, and vertical, $n_b$:
\begin{align*}
n_a&=\Delta\mathbf{r}_{||} \cos{\alpha} - \Delta\mathbf{r}_\perp \sin{\alpha}\\
n_b&=\Delta\mathbf{r}_\perp \cos{\alpha} + \Delta\mathbf{r}_{||} \sin{\alpha}.
\end{align*}
The free energy is just $-T$ times the entropy of paths with this mixture of the two kinds of edges:
\begin{align}\label{mixentropy}
F(\Delta \mathbf{r}_{||},\Delta \mathbf{r}_\perp)&=-\beta^{-1}\log{\binom{n_a+n_b}{n_a}}\\
&\sim -\beta^{-1}\left(
n_a\log{\left(\frac{n_a+n_b}{n_a}\right)}  +n_b\log{\left(\frac{n_a+n_b}{n_b}\right)}\right).\notag
\end{align}
In the second expression we used Stirling's formula for the factorials, since our domain can still have many edges and still be microscopic. This free energy has a regular Taylor series in $\Delta \mathbf{r}_\perp$, the local deviation of the slope from the average slope of the macroscopic system. To second order we obtain, 
\begin{equation}\label{microfree}
F(\Delta \mathbf{r}_{||},\Delta \mathbf{r}_\perp)=\Delta \mathbf{r}_{||}\left( f_0+f_1(\Delta\mathbf{r}_\perp/\Delta \mathbf{r}_{||}) +\frac{1}{2}f_2 (\Delta\mathbf{r}_\perp/\Delta \mathbf{r}_{||})^2+\cdots\right),
\end{equation}
where only the coefficient of the quadratic term will have any bearing on long range order:
\begin{equation*}
f_2=\left(\beta (\sin{\alpha}+\cos{\alpha})\sin{\alpha}\cos{\alpha}\right)^{-1}.
\end{equation*}
The positivity of $f_2$ ($0<\alpha<\pi/2$) can be traced to the convexity of the mixture-entropy \eqref{mixentropy}. We see that the structure of \eqref{microfree} is the volume of the domain in the $\mathbf{r}_{||}$ subspace times an expansion in the gradient of the macroscopic function $\mathbf{r}_\perp(\mathbf{r}_{||})$: 
\begin{equation}\label{freeenergyfunc}
F(\mathbf{r}_\perp)=\int d\mathbf{r}_{||}\left(f_0+f_1\,\partial_{||}\mathbf{r}_\perp  +\frac{1}{2}f_2(\partial_{||}\mathbf{r}_\perp)^2+\cdots\right).
\end{equation}

As in the case of hard spheres, we treat the free energy functional \eqref{freeenergyfunc} as a Hamiltonian for the macroscopic degrees of freedom, the function $\mathbf{r}_\perp(\mathbf{r}_{||})$. For periodic boundary values the linear gradient term integrates to zero and we obtain another instance of the height model. The main result for the height model is that the heights have a distribution $p(\mathbf{r}_\perp)$ above every $\mathbf{r}_{||}$ whose width grows with the system size only when the dimension $D$ is one or two. Previously we argued that the structure factor \eqref{quasistrucfac} would scale with the system size (Bragg peak behavior) when the distribution of $\mathbf{r}_\perp$ was bounded. While this is not the case for our $D=1$ solid of two lengths, analogues of this model in three dimensions produce Bragg peaks because their $\mathbf{r}_\perp$ distribution is independent of system size.

\begin{figure}
\includegraphics{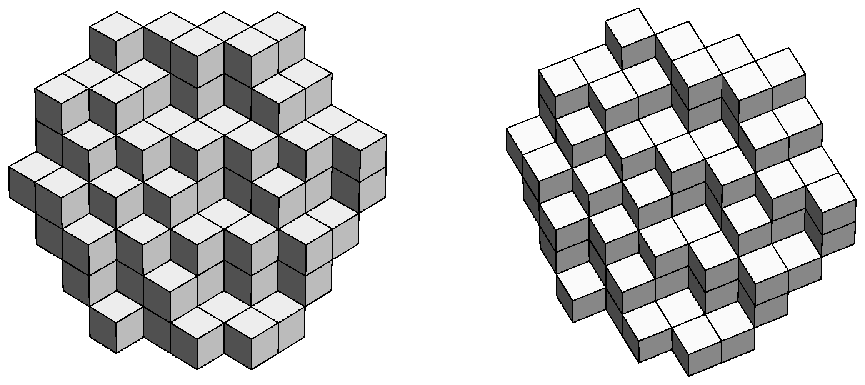}
\caption{Tilings of the plane by three kinds of parallelogram are projections of surfaces formed from the three kinds of square facets in a cubic lattice. In the most symmetrical projection, on the left, the parallelograms are congruent $60^\circ$-rhombi, whose vertices always lie on a hexagonal lattice. }
\label{}
\end{figure}

Figure 7 shows a tiling model for the marginal case, $D=2$. By projecting square facets forming a surface in the cubic lattice we generate tilings of the plane, $\mathbf{r}_{||}$, by three kinds of parallelograms. In the most symmetrical case, when the orthogonal height space $\mathbf{r}_\perp$ is parallel to a 3-fold axis of the cubic lattice, the parallelograms are congruent $60^\circ$-rhombi. This geometry, for a statistical model in the plane, would appear to be the most natural by having just a single structural motif. In addition to lacking perfect long range order as a result of a logarithmically growing height distribution, this model also suffers from the defect that the height distribution cannot be inferred from the rhombus vertices. All heights collapse to a simple hexagonal lattice and the formation of Bragg peaks is trivial. One could avoid this with incommensurate projection spaces, as in the right panel of Figure 7, but then the corresponding tiles are considerably less natural.

The minimum dimension for long range order ($D=3$), incommensurate projection subspaces, and high symmetry, all come together in the six dimensional hyper-cubic lattice \cite{Kramer}. A tiling of space by two ``golden rhombohedra" is constructed in direct analogy with the $D=1$ and $D=2$ constructions described above. Figure 8 shows how exactly two tile shapes emerge when the three dimensional $\mathbf{r}_{||}$ subspace is chosen so the six edges of the hyper-cubic lattice project to the six 5-fold axes of the regular icosahedron. Because the orthogonal $\mathbf{r}_{\perp}$ subspace also has three dimensions, three ``height" variables are required to describe how the three dimensional hyper-surface of face-connected 3-facets meanders through the hyper-cubic lattice. As in the one dimensional model analyzed earlier, the hyper-surface $\mathbf{r}_\perp(\mathbf{r}_{||})$ on macroscopic scales has a squared-gradient free energy density in the most random scenario microscopically, where all tile arrangements have equal probabilities. And because the heights $\mathbf{r}_\perp$ have long range order in three dimensional squared-gradient models, diffraction from such a \textit{random tiling} structure will exhibit Bragg peaks, much like a crystal \cite{Elser2}. The icosahedron-symmetric positions $\mathbf{q}_{||}$ of the Bragg peaks (projected from the hyper-cubic dual lattice vectors $\mathbf{q}$) place this structure in the quasicrystal class.

\begin{figure}[t!]
\includegraphics{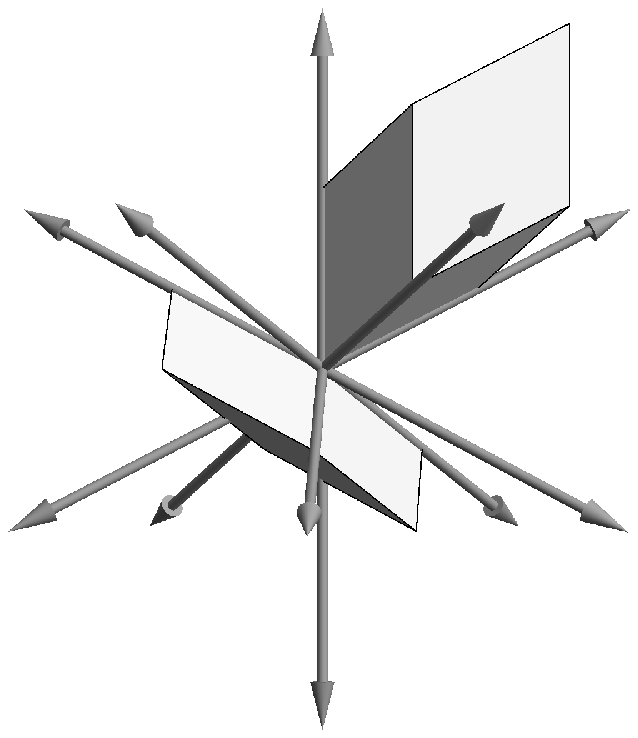}
\caption{The 3-facets of the six dimensional hyper-cubic lattice project to ``golden rhombohedra" when the projection subspace is invariant with respect to the icosahedral group. Edges of the lattice project to the six 5-fold axes of the regular icosahedron, and all triples of distinct edges form a rhombohedron congruent to one of the two shown.}
\label{}
\end{figure}

\newpage

\section*{Problems for study}

\subsection{Free volumes of nearly close-packed spheres}

Compare the free volumes of spheres in the limit of close packing for the fcc and hcp structures. As with disks, fix the surrounding spheres at their average position and approximate the spherical surfaces of constraint by planes. Use symmetry to argue the two free volumes are equal and thereby avoid having to calculate them.

\subsection{Isotropic elasticity of hexagonally packed disks}

The two shear degrees of freedom of a material in two dimensions are described by the distortion matrix
\begin{equation*}
\mathbf{A}=\left[
\begin{array}{cc}
a_1&a_2\\
a_2&-a_1
\end{array}
\right].
\end{equation*}
Rotating the material by $60^\circ$ gives a different distortion
\begin{equation*}
\mathbf{A}=\left[
\begin{array}{cc}
a'_1&a'_2\\
a'_2&-a'_1
\end{array}
\right].
\end{equation*}
Determine $a'_1$ and $a'_2$ in terms of $a_1$ and $a_2$ and use this result to argue that if the shear elastic free energy
of hexagonally packed disks is $\kappa_1 a_1^2+\kappa_2 a_2^2$ for general $a_1$ and $a_2$, then $\kappa_1=\kappa_2$.
Hexagonally packed disks are thus elastically isotropic: the free energy makes no reference to the crystal axes.

\subsection{Height model}

Fill in all the missing steps in the analysis of the height model.

\subsection{Ising model with many ground states}

In some solids the only degrees of freedom that have significant entropy at low temperature are the electron magnetic moments, or \textit{spins}. Often the Hamiltonian of such solids can be modeled by spin variables that take two values, $s=\pm1$, corresponding to the magnetic moment along a particular axis in units of $\hbar/2$. Consider a two dimensional solid where the spins are arranged on a hexagonal lattice and have the Ising Hamiltonian
\begin{equation*}
H=J\sum_{(i j)} s_i s_j,
\end{equation*}
where the sum is over all adjacent pairs of spins and the coupling $J$ is positive (antiferromagnetism). To get the lowest possible energy, or ground state, we would like adjacent pairs of spins to have opposite sign. But this is impossible, because the adjacency graph for the spins is not bipartite. Show that this system has in fact many ground states, and that these are in 2-to-1 correspondence with the $60^\circ$-rhombus tilings of the plane.









\end{document}